\documentclass[aps,prl,onecolumn,showpacs,groupedaddress,notitlepage,longbibliography]{revtex4-1}
\usepackage[centertags]{amsmath}
\usepackage{graphicx}
\usepackage{amsfonts}
\usepackage{epstopdf}
\usepackage{color}

\begin{document}
\title{Phase-suppressed hydrodynamics of solitons on constant-background plane wave}

\author{A. Chabchoub$^{1,2,3,\ast}$, T. Waseda$^3$, M. Klein$^4$, S. Trillo$^5$, and M. Onorato$^{6,7}$}
\affiliation{$^1$ Centre for Wind, Waves and Water, School of Civil Engineering, The University of Sydney, Sydney, New South Wales 2006, Australia} 
\email{amin.chabchoub@sydney.edu.au}
\affiliation{$^2$ Marine Studies Institute, The University of Sydney, Sydney, New South Wales 2006, Australia}
\affiliation{$^3$ Department of Ocean Technology Policy and Environment, Graduate School of Frontier Sciences, The University of Tokyo, Kashiwa, Chiba 277-8563, Japan}
\affiliation{$^4$ Institute for Structural Dynamics, Hamburg University of Technology, 21073 Hamburg, Germany}
\affiliation{$^5$ Department of Engineering, University of Ferrara, 44122 Ferrara, Italy}
\affiliation{$^6$ Dipartimento di Fisica, Universit\`a degli Studi di Torino, 10125 Torino, Italy}
\affiliation{$^7$ Istituto Nazionale di Fisica Nucleare, INFN, Sezione di Torino, 10125 Torino, Italy}

\begin{abstract}
Soliton and breather solutions of the nonlinear Schr\"odinger equation (NLSE) are known to model localized structures in nonlinear dispersive media such as on the water surface. One of the conditions for an accurate propagation of such exact solutions is the proper generation of the exact initial phase-shift profile in the carrier wave, as defined by the NLSE envelope at a specific time or location. Here, we show experimentally the significance of such initial exact phase excitation during the hydrodynamic propagation of localized envelope solitons and breathers, which modulate a plane wave of constant amplitude (finite background). Using the example of stationary black solitons in intermediate water depth and pulsating Peregrine breathers in deep-water, we show how these localized envelopes disintegrate while they evolve over a significant long distance when the initial phase shift is zero. By setting the envelope phases to zero, the dark solitons will disintegrate into two gray-type solitons and dispersive elements. In the case of the doubly-localized Peregrine breather the maximal amplification is considerably retarded; however locally, the shape of the maximal focused wave measured together with the respective signature phase-shift are almost identical to the exact analytical Peregrine characterization at its maximal compression location. The experiments, conducted in large-scaled shallow-water as well as deep-water wave facilities, are in very good agreement with NLSE simulations for all cases. 
\end{abstract}
\maketitle
\section{Introduction}
Wave propagation in nonlinear dispersive media are known to be modeled by weakly nonlinear evolution equations such as the nonlinear Schr\"odinger equation (NLSE) \cite{benney1967propagation,zakharov1968stability}. Even though the NLSE is restricted in taking into account only weak nonlinearities of the wave field  and narrow-band processes, several laboratory studies have confirmed its validity to predict the dynamics of stationary and pulsating coherent structures in optics, hydrodynamics and plasma \cite{yuen1982nonlinear,hasegawa1989optical,dauxois2006physics,kibler2010peregrine,chabchoub2011rogue,bailung2011observation}. One remarkable feature of the hydrodynamic NLSE is that it takes into account the correct ratio of group and phase velocity even when higher-order effects are at play \cite{henderson1999unsteady}, 
which are captured in the modified NLSE framework \cite{dysthe1979note,goullet2011numerical}.   

Indeed, the vast and manifold families of exact NLSE solutions allow to quantitatively study the dynamics of localized structures, especially, within an experimental framework \cite{akhmediev1997solitons,onorato2013roguereport,dudley2019rogue}, since they provide an exact parametrization of the wave field in time and space.
In particular, the time-NLSE provides a convenient framework where the evolution of the wave packet is described in space along the longitudinal propagation coordinate, with the temporal coordinate playing the role of transverse variable.
When modelling a localized wave packet on a finite background (i.e. a carrier wave of constant amplitude and single frequency), 
an exact solution of the time-NLSE determines and applies a specific temporal transverse phase-shift profile with respect to the background wave, which can evolve or remain constant. This fundamental feature is an essential attribute to ensure coherence. Without the correct temporal phase-shift in the initial and launching condition of the NLSE solution, the envelope is expected to diverge from the anticipated trajectory, resulting in the disintegration into several {\it similar} or different localized structures. 

NLSE solutions with finite background differ in the defocusing and focusing regimes, respectively, which describe water wave propagation at different water depths. In the defocusing regime the NLSE supports one-soliton solutions in the form of stationary dark solitons \cite{shabat1973exact} or multi-soliton dark solutions \cite{Fratalocchi2008}. The simplest one-soliton solutions are characterised by a single parameter that fixes the maximal phase-shift along the envelope profile ranging from 0 to $\pi$, which in turn is related to the darkness of the envelope and to its velocity with respect to the reference group velocity \cite{chabchoub2014gray}.
The limiting case corresponds to the black solution which has a characteristic shift of $\pi$, zero velocity, and total darkness at the dip. This shift in the carrier is expected to remain constant throughout the whole propagation in time and space. 


Conversely, in the focusing case, the solutions are bright and pulsating and have obviously the property of exhibiting a phase profile which varies with the focusing of the envelope. 
The most fundamental solution is the Peregrine breather, which describes the modulation instability for the case of infinite modulation period \cite{peregrine1983water}.
When modelled in the framework of the time-NLSE, the Peregrine breather has a temporal phase profile with maximum shift between the background and the peak elevation that ranges from 0 to $\pi$ upon evolution, where the maximum value of $\pi$ occurs across the two transverse zeros of the envelope, and is achieved exactly when the wave packet reaches its maximal focusing point \cite{kedziora2013phase,xu2019phase}.
This abrupt temporal $\pi$-shift at the focus point is the most distinctive feature of Peregrine solitons that we specifically address here.


The general aim of this paper is to investigate numerically and experimentally the significance of the initial transverse temporal phase profile of the launched soliton and breather hydrodynamics using the black soliton and the Peregrine breather as references. Considering a background field with input negative (dip) envelope modulation in the defocusing or positive (bump) modulation in the focusing regime,  respectively, we show that the suppression of the appropriate phase-shift would engender the disintegration of the localization.
All reported experimental results are in excellent agreement with numerical NLSE simulations. We also discuss the long-term propagation and the universal feature of solitons and breathers deprived from their distinctive phase setting. 

\section{Localized Envelopes and Experimental Set-Up} 

Nonlinear waves in intermediate water depth as well as in deep-water can be described by the defocusing and the focusing time-NLSE, respectively. In dimensional form, this evolution equation reads \cite{hasimoto1972nonlinear}
\begin{eqnarray}
-\operatorname{i}\left(\frac{\displaystyle\partial\psi}{\displaystyle \partial
x}+\beta_1\frac{\displaystyle\partial\psi}{\displaystyle \partial
t}\right)+\dfrac{1}{2}\beta_2\frac{\displaystyle\partial^2\psi}{\displaystyle \partial
t^2}+\gamma\left|\psi\right|^2\psi=0, \label{dnls}
\end{eqnarray}
where: 
\begin{align}
\beta_1=&\dfrac{1}{c_g},\\
\beta_2=&-\dfrac{1}{c_g^3}\dfrac{\partial^2\omega}{\partial k^2},\\
\gamma=&\dfrac{\omega k^2}{16c_g\sinh^4\left(kh\right)}\left(\cosh\left(4kh\right)+8-2\tanh^2\left(kh\right)\right)\nonumber\\
&-\dfrac{\omega}{2\sinh^2(2kh)}\dfrac{\left(2\omega\cosh^2\left(kh\right)+kc_g\right)^2}{c_g\left(gh-c_g^2\right)}.
\end{align}
Here, $g$ denotes the gravitational acceleration, $h$ is the water depth,  $k$ is the wave number of the carrier wave, while the dispersion relation reads $\omega=\sqrt{gk\tanh{kh}}$ and $c_g=\dfrac{\partial\omega}{\partial k}$ is the group velocity of the  wave packets. When $kh<1.363$, i.e. shallow or intermediate water depth, $\beta_2\gamma<0$ the NLSE is known to admits a family of dark one-soliton solutions \cite{shabat1973exact,chabchoub2014gray}, which are characterized by a negative envelope dip.
 
One limiting case of the family of dark solitons, which locally diminishes the amplitude of the carrier to zero, is referred to as black soliton and has the following simple form for a given normalized background amplitude 
\begin{eqnarray}
\psi_B\left(x,t\right)=&\tanh\left[\sqrt{\dfrac{-\gamma}{\beta_2}}\left(t-\dfrac{x}{c_g}\right)\right]\exp\left(-\operatorname{i}\gamma x\right).
\end{eqnarray}
This fundamental solution has been observed in a wide range of nonlinear dispersive media, for instance in optics \cite{weiner1988experimental}, Bose-Einstein condensates \cite{frantzeskakis2010dark}, plasma \cite{shukla2006formation} and recently also in hydrodynamics \cite{chabchoub2013experimental}. 

In deep-water, when $kh>1.363$, $\beta_2\gamma>0$ and, besides the propagation of stationary envelope soliton packet, the NLSE can also accurately describe the modulation instability dynamics \cite{osborne2010nonlinear}. Indeed, the family of Akhmediev breathers describe the Stokes waves' instability for each case of unstable modulation frequency \cite{akhmediev1985generation,akhmediev1987exact}. The limiting case of zero modulation frequency is analytically described by the Peregrine breather \cite{peregrine1983water}, parametrized as
\begin{eqnarray}
\psi_P\left(x,t\right)=\left(-1+\dfrac{4-8\operatorname{i}\gamma x}{1+4\dfrac{\gamma}{\beta_2}\left(t-\dfrac{x}{c_g}\right)^2+4\gamma^2 x^2}\right)\exp \left(-\operatorname{i}\gamma x\right). 
\label{Peregrine_par}
\end{eqnarray}
Note that the maximal wave focusing occurs at $x=0$ in this parametrization.

The corresponding dimensional spatiotemporal surface elevation, taking into account the second-order Stokes correction, in both, focusing and defocusing cases, is modelled by
\begin{eqnarray}
\eta\left(x,t\right)=\textnormal{Re}\left(\psi\left(x,t\right)\exp\left[\textnormal{i}\left(kx-\omega t\right)\right]\right.+\dfrac{1}{2}k\psi^2\left(x,t\right)\exp\left[2\textnormal{i}\left(kx-\omega t\right)\right]). 
\label{se}
\end{eqnarray}
Equation (\ref{se}) can be used to determine the experimental boundary conditions for the hydrodynamic experiments. These NLSE solutions can be represented as $\psi\left(x,t\right)=A\left(x,t\right)\exp\left[\operatorname{i}\varphi\left(x,t\right)\right]$ \cite{osborne2010nonlinear} and hold a phase-shift in the complex envelope, which is also transmitted to the corresponding water surface elevation signals. That said, each exact NLSE solution assigns a local phase dynamics, which is determined by the parametrization of the solution. In order to physically observe such solutions, it is mandatory to take these phase-shifts into consideration in defining the initial 
conditions as prescribed by Eq. (\ref{se}). In order to ignore this phase-shift in the initial condition, as designated by the exact NLSE solution while keeping the same initial envelope configuration and geometry, we have to simply replace, at some specific location $x$,  $\psi\left(x,t\right)$ by $\left|\psi\left(x,t\right)\right|$ in Eq. (\ref{se}) 
\begin{eqnarray}
\eta\left(x,t\right)=\textnormal{Re}\left(\left|\psi\left(x,t\right)\right|\exp\left[\textnormal{i}\left(kx-\omega t\right)\right]\right.+\dfrac{1}{2}k\left|\psi\left(x,t\right)\right|^2\exp\left[2\textnormal{i}\left(kx-\omega t\right)\right]), 
\label{se2}
\end{eqnarray}which means setting all the phases equal to zero. Consequently, it is expected from weakly nonlinear theory that the envelope dynamics will be significantly different. For instance in the case of the black soliton, the  condition for the stationary wave propagation is going to be violated. We shall investigate this type of wave motion experimentally and numerically in detail together with the Peregrine breather for deep-water conditions. We emphasize that a significant wave propagation in space is required in order to experimentally study the disparity in the wave propagation.  

The experiments have been conducted in two large water wave facilities. The first with water of intermediate depth is installed at the Technical University of Berlin, with dimensions of 110 $\times$ 8 $\times$ 0.4 $ \textnormal{m}^3$ and a piston-type wave maker, whereas the deep-water wave flume installed at The University of Tokyo generates the waves by means of flap-type wave maker while its dimensions are 85 $\times$ 3.5 $\times$ 2.2 $ \textnormal{m}^3$. The remaining configuration of the facility are similar: a number a wave gauges are installed along the wave facilities to measure the evolution of the nonlinear wave field and a wave absorber is installed at the end of the facility in order to avoid the reflection of the waves.

\section{Experimental and numerical results}
\subsection{Dark solitons }
We will first start with the description of the experiments in the defocusing regime. We recall that in order to generate dark solitons, we have to satisfy the hydrodynamic dimensionless depth condition $kh<1.363$ \cite{chabchoub2013experimental}. 

\begin{center}
\begin{figure}[h]
\begin{tabular}{cc}
\includegraphics[width=.48\textwidth]{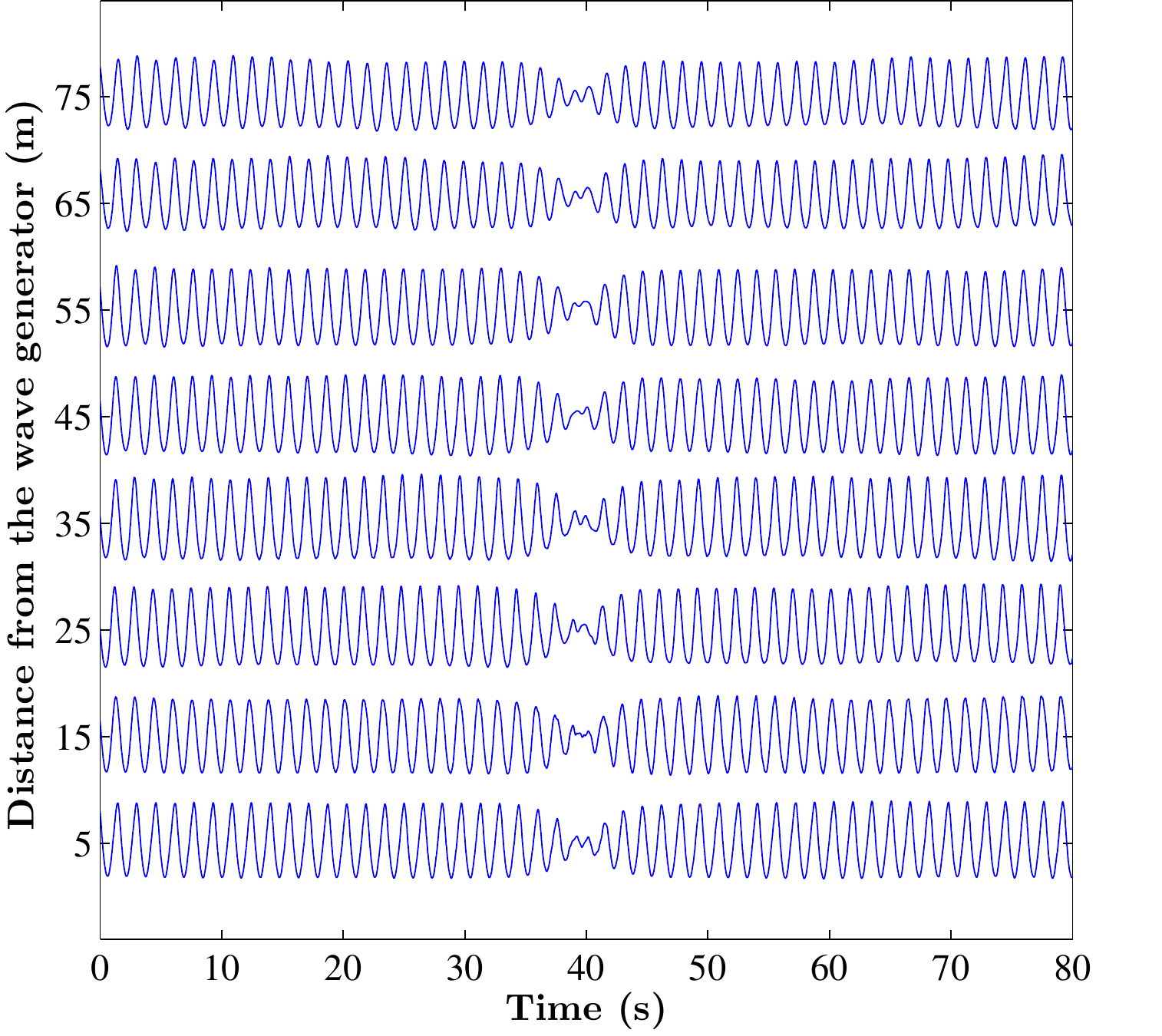}&
\includegraphics[width=.48\textwidth]{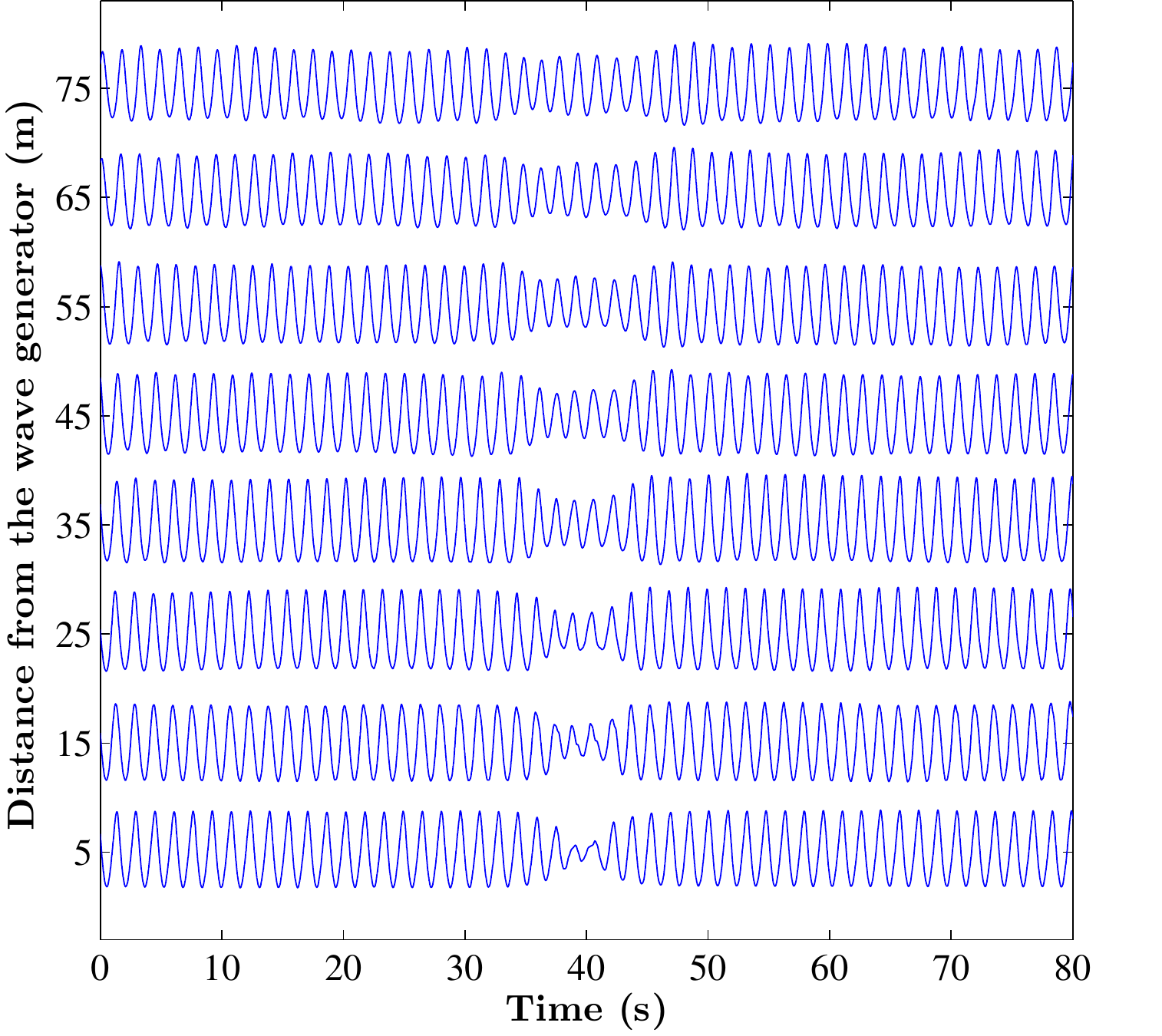}\\
\end{tabular}
\caption{Left: Propagation of black soliton, starting with an exact initial condition, as described from the exact NLSE framework. The amplitude of the carrier is $a=0.036$ m, the steepness $\varepsilon=k a=0.08$, the still water depth is $h=0.4$ m corresponding to a dimensionless depth $kh=0.9$. Right: Propagation of a similiar initial localized structure having the carrier parameter as in left panel, however, suppressing the characteristic black soliton phase-shift of $\pi$ in the initial conditions.
}
\label{fig1}
\end{figure}
\end{center}

A first set of experiments is carried out with carrier parameters determined by the amplitude $a=0.036$ m, a steepness of $\varepsilon=0.08$ for a dimensionless depth of $kh=0.9$. Figure \ref{fig1} shows the evolution of the black soliton wave (left panel) as well as the case of same envelope depression structure without the initial phase-shift of $\pi$ \cite{chabchoub2014gray} (right panel), both propagating over a considerable propagation distance of 75 m. Note that at the first gauge, i.e. at the lowest time series in the Fig.~\ref{fig1}, the surface elevation are almost indistinguishable by eye in the two cases. 
\onecolumngrid
Clearly, we can observe a very {\it clean} stationary propagation of the black soliton, as expected form NLSE predictions and as already reported 
by means of a different facility \cite{chabchoub2013experimental}. 
In the latter work the wave flume and thus the propagation distance was significantly shorter as in these reported tests. Note, that the present one is to date the longest propagation of a black soliton reported in water waves and the result proves the robustness of such localized structures during their evolution. The right panel in Fig.~\ref{fig1} shows the result of the same type of experiment, however, when the $\pi$-phase shift is ignored in the initial conditions the dynamics evidently differs. Indeed, a visible distortion of the wave field is noticed and the initial envelope dip in the dark wave envelope does not remain localized, but rather leads to fission into two shallower envelope dips that separate from each other with definite (opposite) group velocities, since the coherence is distorted and dispersive effects become more significant. This suggests that the initial profile is embracing in this case two gray solitons and weaker dispersive waves. It is expected that this can be detected using advanced inverse scattering eigenvalue analysis \cite{Fratalocchi2008,randoux2018nonlinear,osborne2019breather}. The latter envelope distortions is discussed below in more detail with reference to Fig. \ref{fig4}.

\onecolumngrid
\begin{center}
\begin{figure}[h]
\centering
\begin{tabular}{cc}
\includegraphics[width=.48\textwidth]{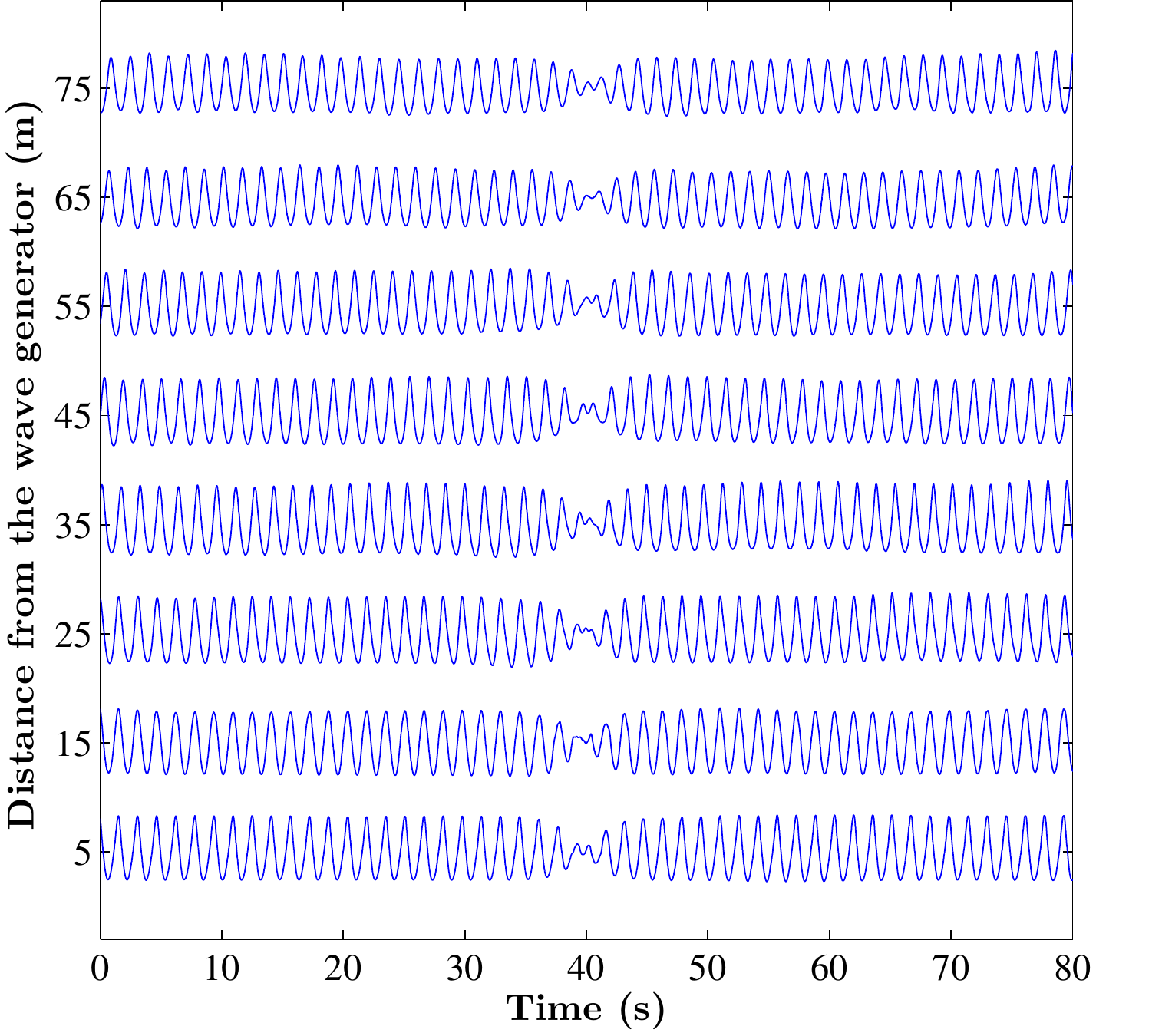}&
\includegraphics[width=.48\textwidth]{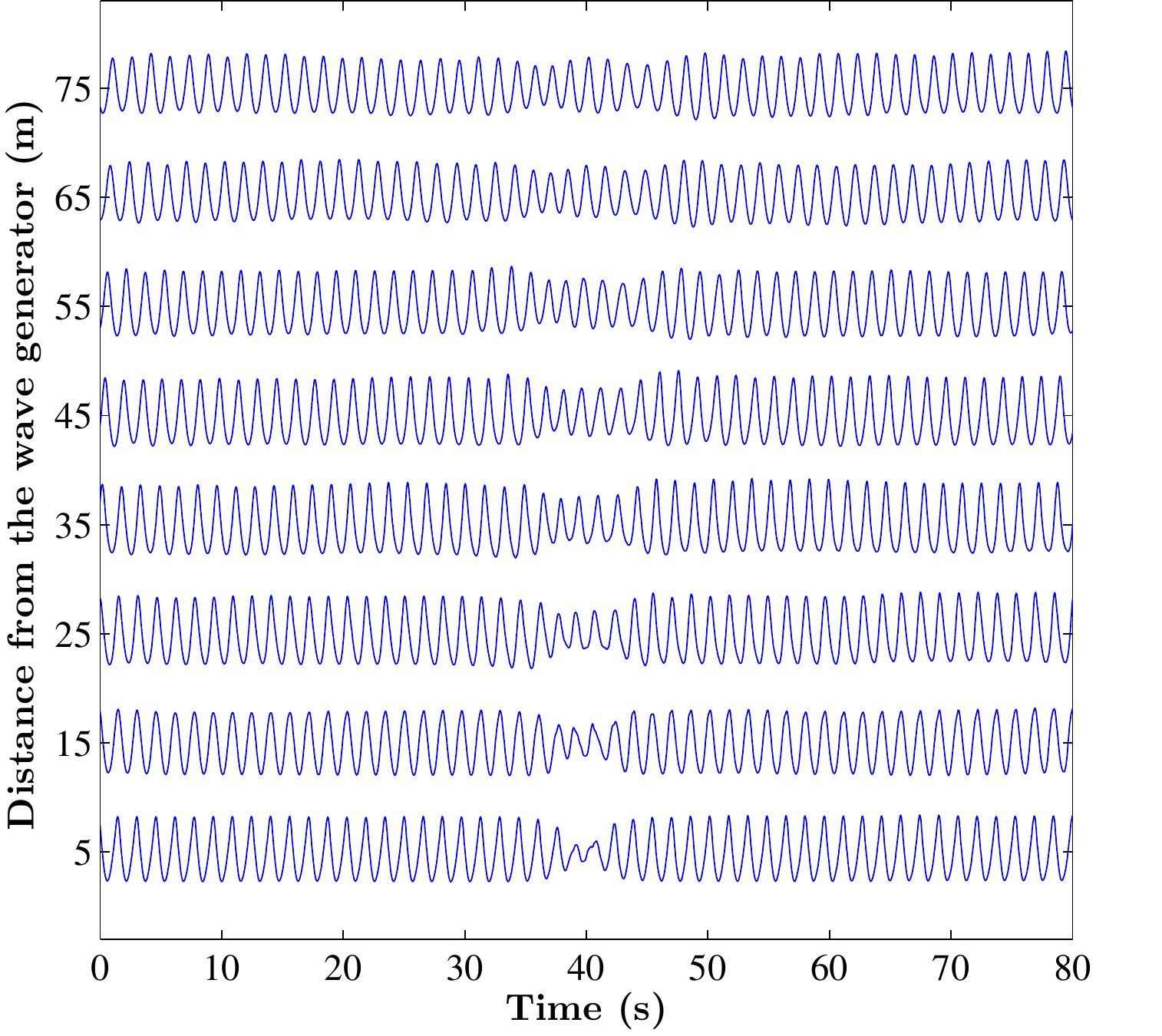}\\
\end{tabular}
\caption{Left: Propagation of black soliton, starting with an exact initial condition, as described from the exact NLSE framework. 
Right: Propagation of a similar initial localized structure having the carrier parameter as in (Left), however, ignoring the characteristic black soliton phase-shift of $\pi$ in the initial conditions. 
Compared with data in Fig.~\ref{fig1}, the water depth is the same ($h=0.4$ m,  $kh=0.9$), but the carrier amplitude is larger, $a=0.045$ m, corresponding to higher steepness $\varepsilon=ak=0.1$.
}
\label{fig2}
\end{figure}
\end{center}

A second example of the same type of experiments is illustrated by choosing different carrier parameters. In this example, we slightly modify the degree of nonlinearity by increasing the amplitude of the background to $a=0.045$ m so that the steepness becomes $\varepsilon=0.10$. The experimental results are depicted in Fig.~\ref{fig2}.
As in Fig.~\ref{fig1}, we observe again an ideal propagation of the black soliton for the initial conditions dictated by the NLSE, whereas without correct initial phases, the initial dark zero dip fissures. As we increase the steepness, compared to the first experiment, the degree of nonlinearity is increased as well and the envelope-splitting behaviour is much more pronounced (this will be clearly illustrated in Fig. \ref{fig4}). 
We point out that the observed behavior is consistent with similar observations previously reported in optics \cite{krokel1988twindark,Luther-Davies1992nophase} and Bose-Einstein condensation \cite{Hau2001BEC}.
In order to investigate the nature of the fission in these observations, we also conducted numerical NLSE simulations of these latter four cases, as described above and as shown in Fig.~\ref{fig1} and Fig.~\ref{fig2}. The numerical integration scheme is based on the on the common split-step technique \cite{hardin1973applications,fisher1975numerical}. 
\begin{center}
\begin{figure}[h]
\centering
\begin{tabular}{cc}
\includegraphics[width=.49\columnwidth]{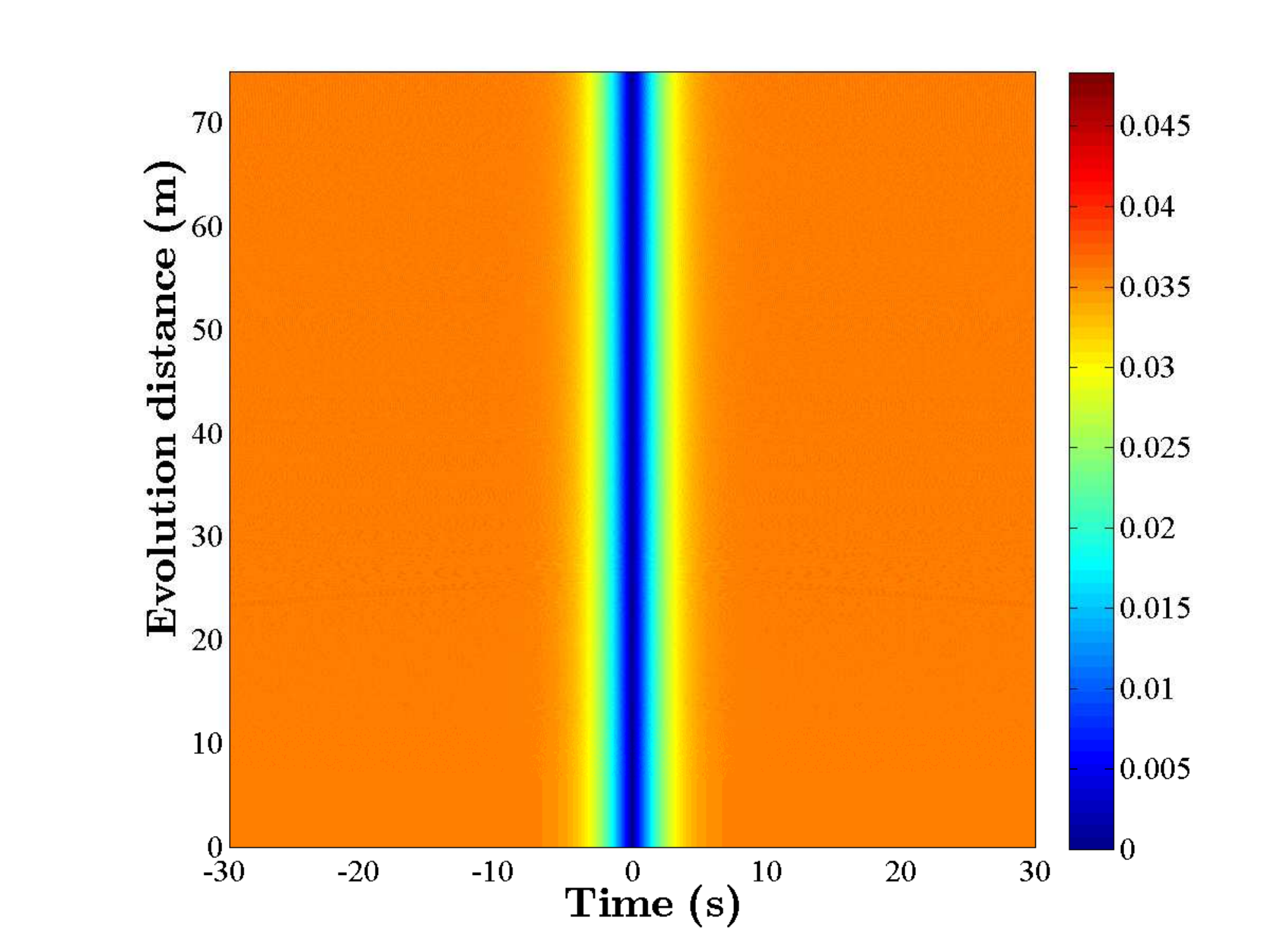}&
\includegraphics[width=.49\columnwidth]{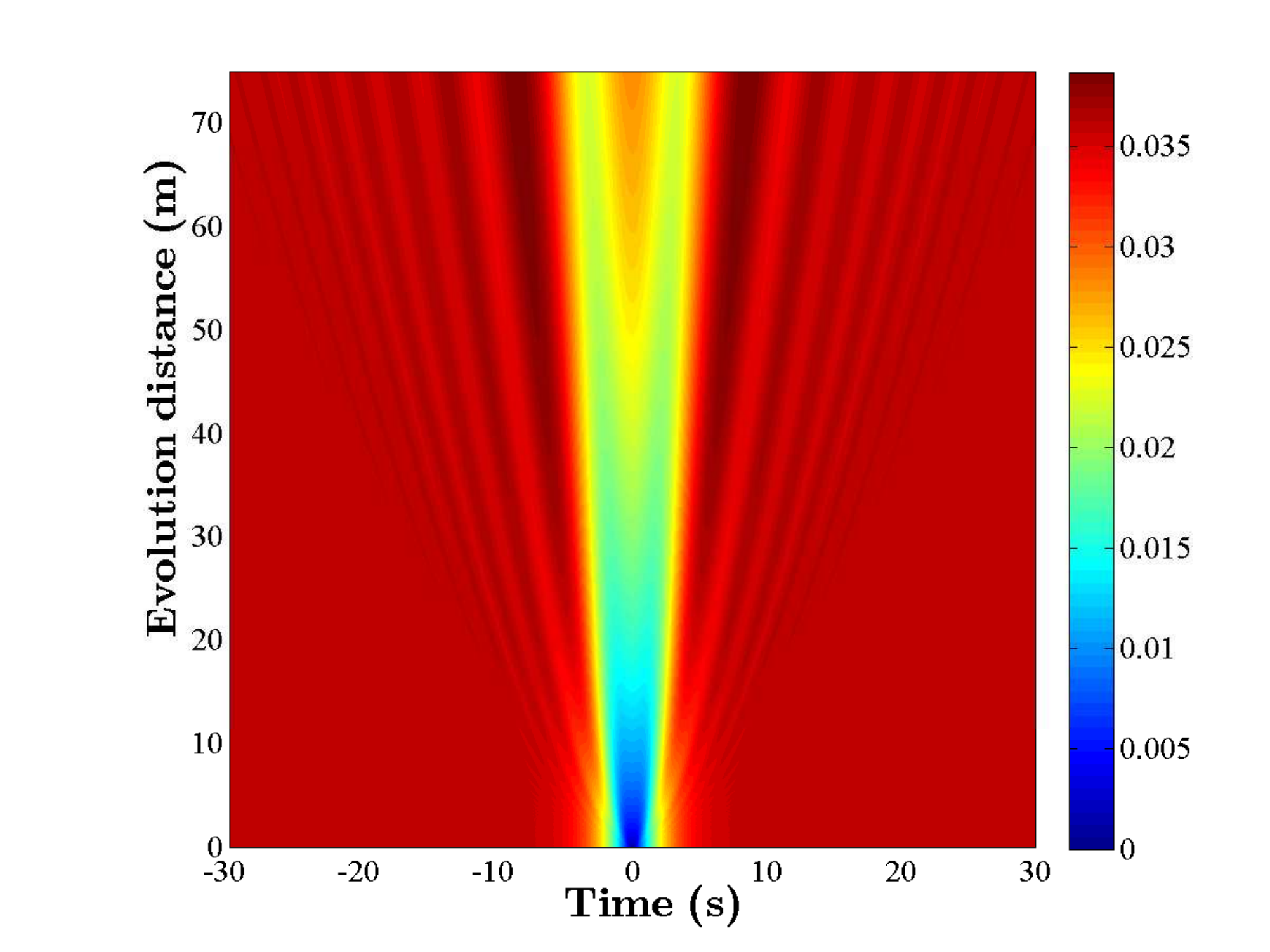}\\
\includegraphics[width=.49\columnwidth]{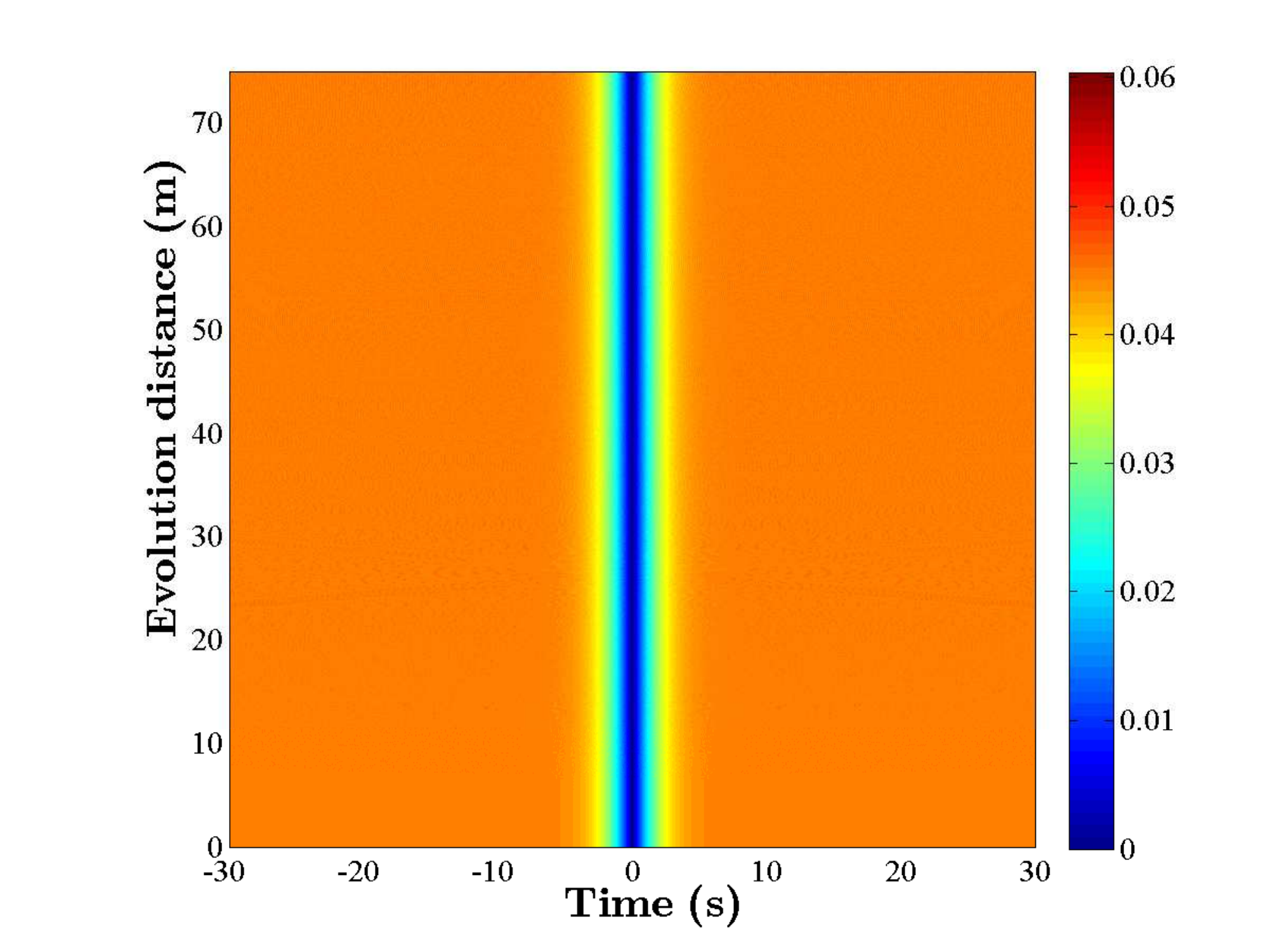}&
\includegraphics[width=.49\columnwidth]{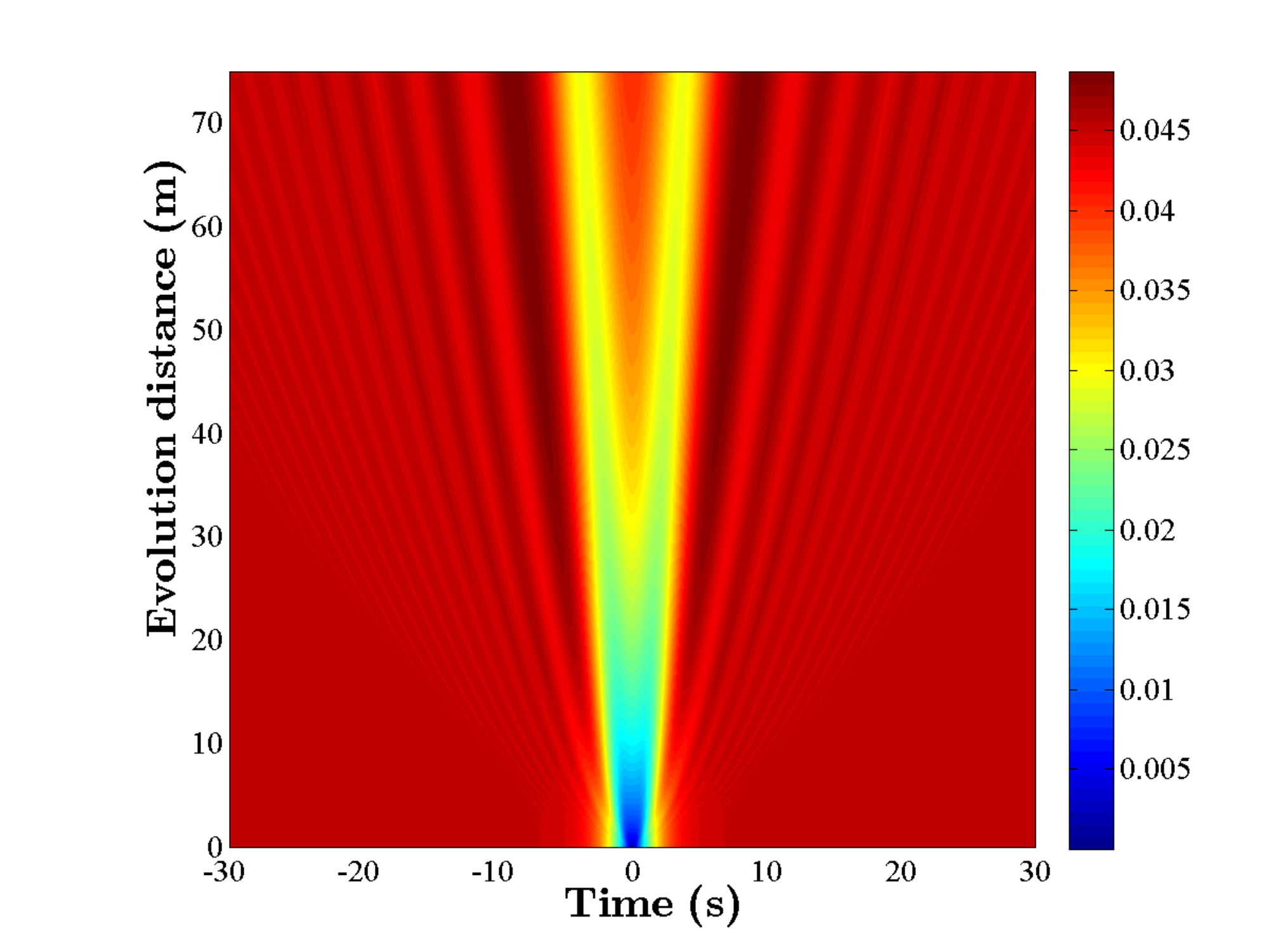}\\
\end{tabular}
\caption{
Numerical NLSE simulation of the experiments in Figs.~\ref{fig1}-\ref{fig2}. The top row is relative to the case shown in Fig.~\ref{fig1} (left: soliton input; right: suppressed phase input).
The bottom row is relative to the case shown in Fig.~\ref{fig2} (left: soliton input; right: suppressed phase input).
}
\label{fig3}
\end{figure}
\end{center}
%
\begin{center}
\begin{figure}[h]
\begin{tabular}{cc}
\includegraphics[width=.48\columnwidth]{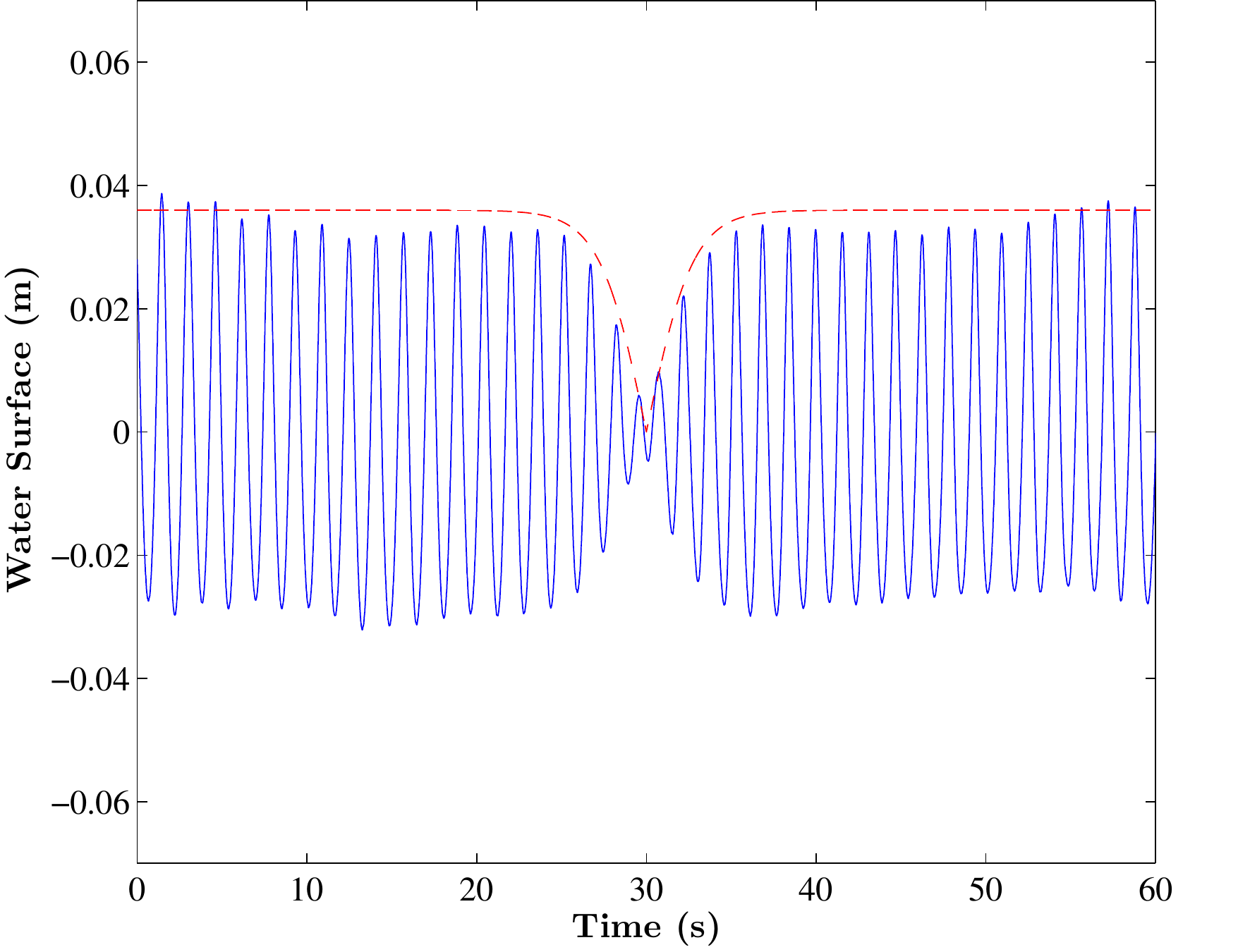}&
\includegraphics[width=.48\columnwidth]{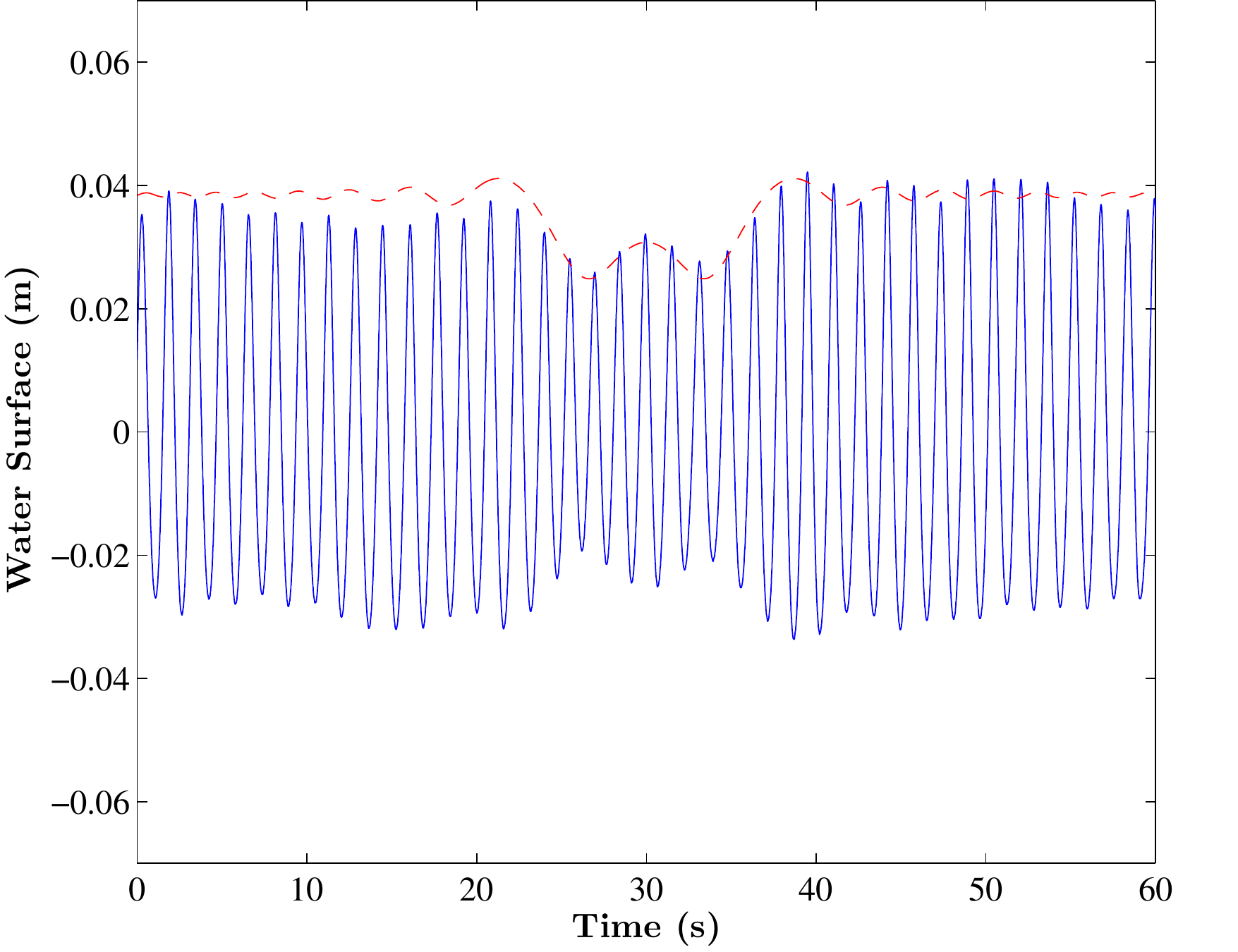}\\
\includegraphics[width=.48\columnwidth]{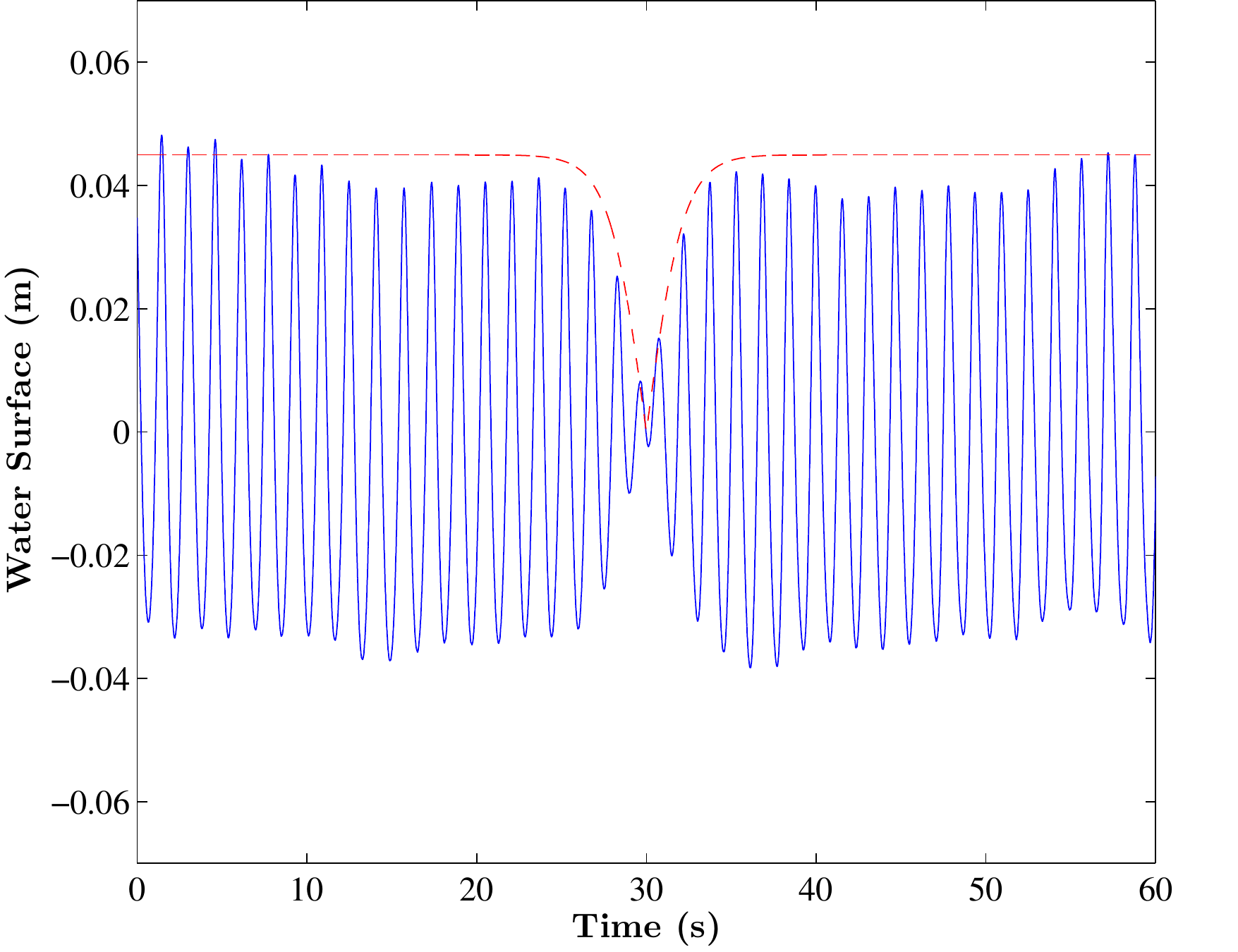}&
\includegraphics[width=.48\columnwidth]{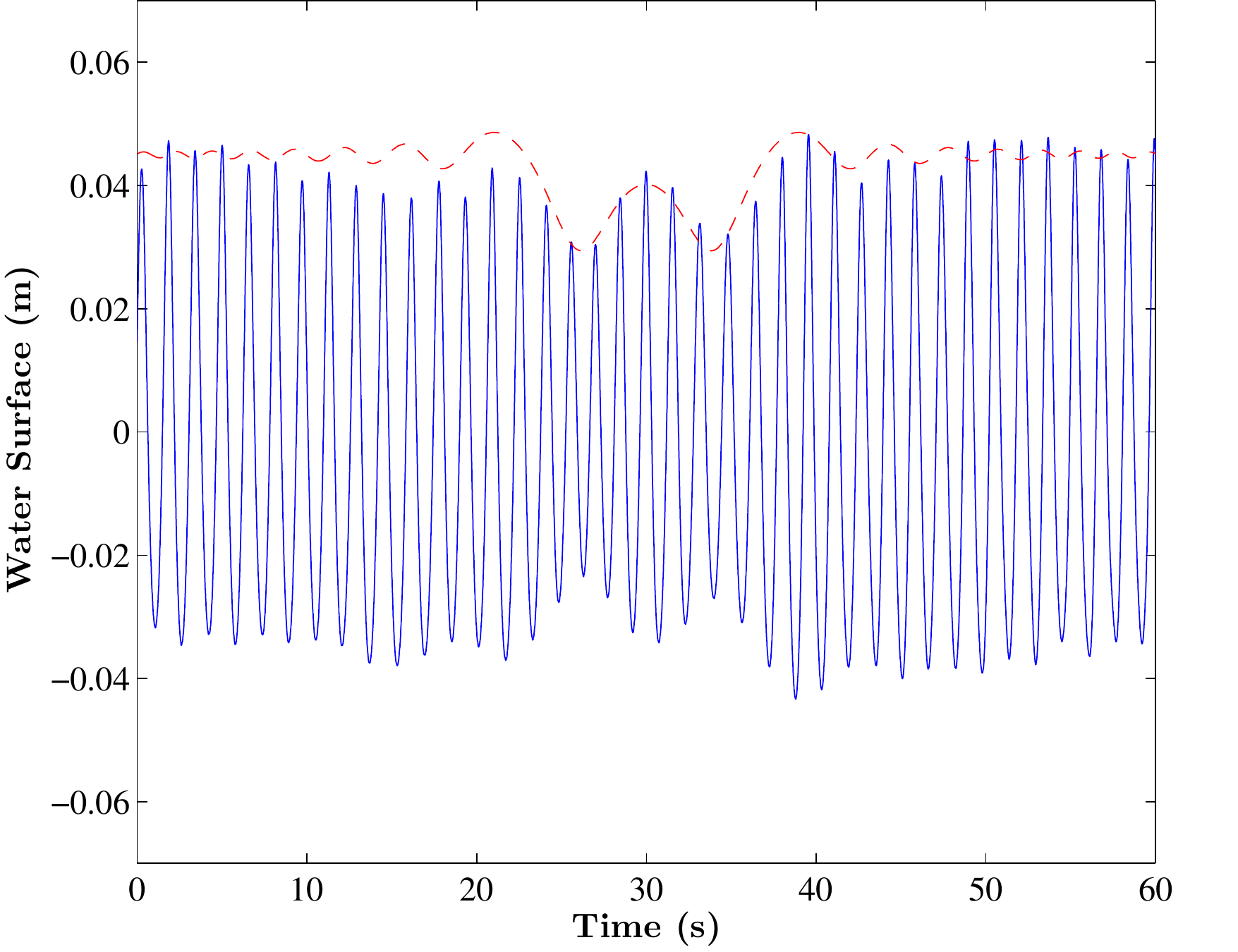}\\
\end{tabular}
\caption{Blue continuous lines: wave tank measurements after 75 m of wave propagation. Red dashed lines: NLSE prediction at the same gauge position. Top Left: Last measurement of Fig. 1 (Left) compared with the exact black soliton solution. Top Right: Last measurement of Fig. 1 (Right) compared with numerical NLSE simulations. Bottom Left: Last measurement of Fig. 2 (Left) compared with corresponding numerical NLSE simulations after 75 m of propagation. Bottom Right: Last measurement of Fig. 2 (Right) compared with corresponding numerical NLSE simulations after 75 m of propagation.}
\label{fig4}
\end{figure}
\end{center}
The simulations of wave envelope profiles as depicted in Fig.~\ref{fig3} (top left panel) and (bottom left panel) have been performed to demonstrate the accuracy of the numerical scheme predicting the evolution of the stationary black soliton and thus, for the sake of accurate interpretation of the numerical results. As can be noticed in such panels,
the initial dark soliton envelopes remain indeed stationary. However, by ignoring the $\pi$-phase-shift of the carrier around the envelope-depression-type localization, we evidently observe the
break-up of the localized envelope zero dip, seemingly into two gray-type structures with additional dispersive waves (ripples) emitted towards the edges of the temporal window.
In this case, the initial envelope cannot stand the absence of phase jump across its zero and breaks into two shallower dips (each with its own phase jump, as we will discuss in the following) that moves away from the initial zero in order to conserve the overall momentum. The excess energy radiated in the form of ripples is caused by the fact that the initial dip with suppressed phase is no longer a soliton solution of the NLSE, and hence its spectral content contains also radiation. 

As next step of the study, we compare the measurements recorded from farthest wave gauge, placed 75 m from the wave maker, with the corresponding envelope prediction, extracted from the NLSE simulations. In other words, the final temporal traces in the four cases shown in Fig.~\ref{fig3} superimposed to the farthest water wave tank observations. These comparison results are shown in Fig.~\ref{fig4}.

Figure \ref{fig4} shows a very good agreement achieved between these farthest measurements, recorded at a distance of 75 m of wave propagation and the corresponding NLSE simulations. This confirms the validity of the NLSE in describing the dynamics of water waves in finite water depth. We also observe a particular agreement for the cases related to the fission, when injecting initial conditions ignoring the $\pi$-phase-shift. We believe that these prediction results are quite remarkable in view of the very long propagation distance trailed while considering that higher-order effects and experimental imperfections are always present. 

Interestingly enough, we point out that the black soliton with suppressed phase shift can break in several pairs (instead of a single pair) of symmetric gray solitons when one enters the semi-classical regime \cite{grimshaw2013rogue,bertola2013universality} which implies a dominant nonlinearity, i.e. for a given width a much larger wave amplitude compared with the soliton amplitude, or for a given amplitude a much larger width compared with soliton width \cite{Moro2014,Hau2001BEC}. This regime, however, is not accessible in our experiment and will be addressed in the future.

\subsection{Peregrine soliton}
\begin{center}
\begin{figure}[h]
\begin{tabular}{cc}
\includegraphics[width=.48\textwidth]{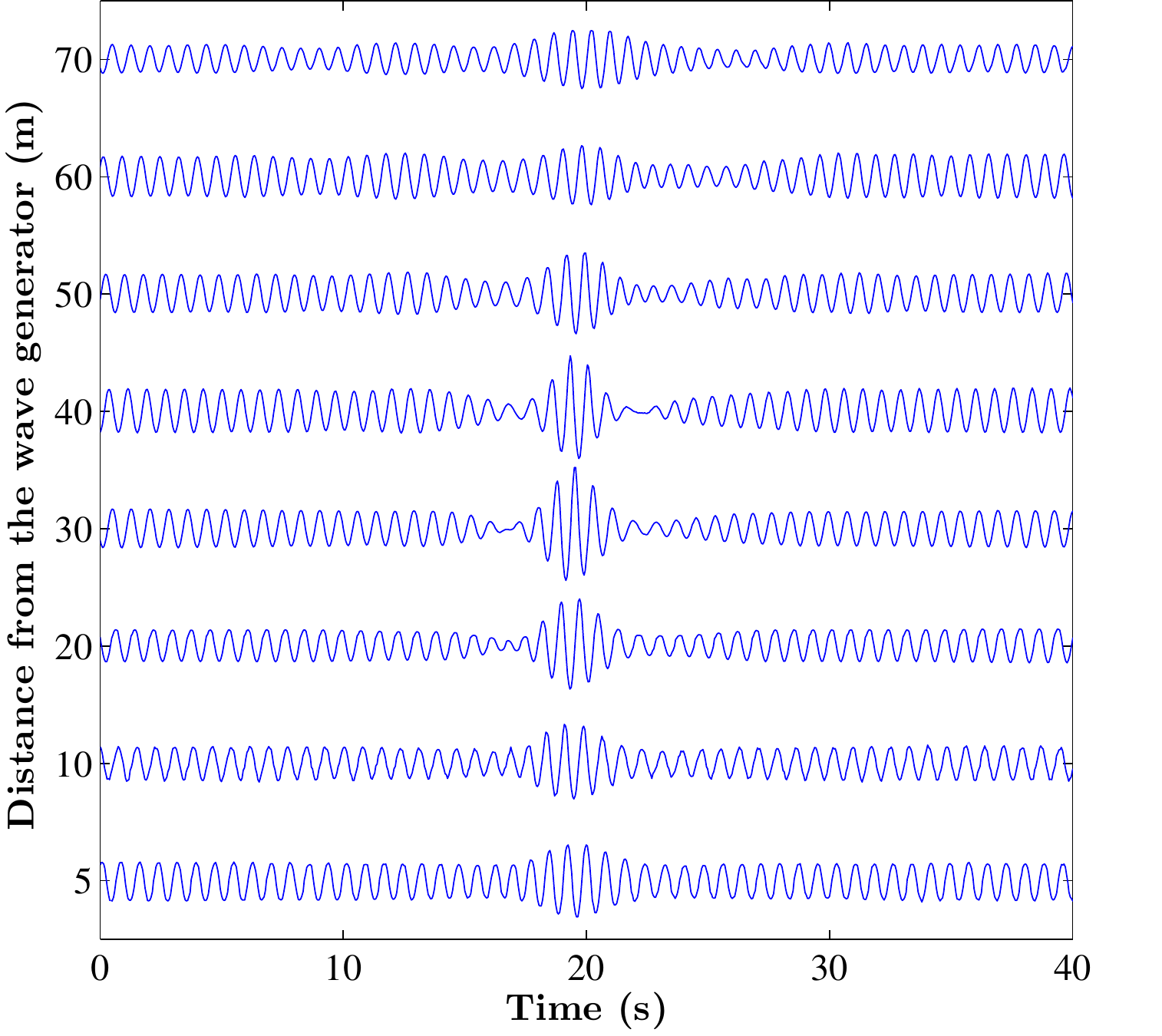}&
\includegraphics[width=.48\textwidth]{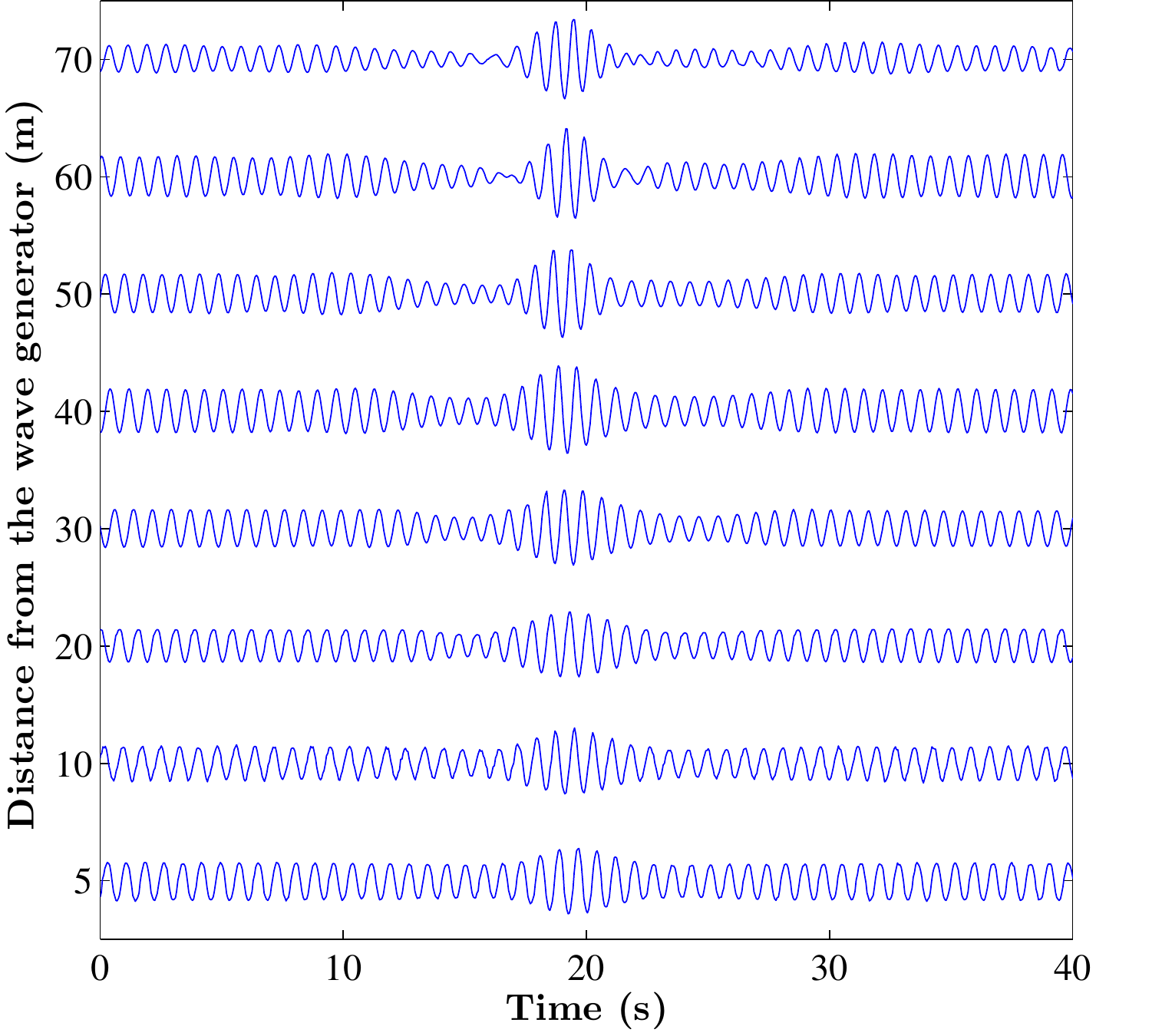}\\
\end{tabular}
\caption{Left: Propagation of Peregrine breather, starting with an exact initial condition as described by Eq. (\ref{Peregrine_par}) at $x=-25$ m. The amplitude of the carrier is $a=0.010$ m, while the steepness is $\varepsilon=0.06$. Right: Propagation of a similar initial breather structure for the same carrier parameters as in left panel, however, suppressing the characteristic Peregrine phase-shift imposed in the exact initial condition (Eq. (\ref{Peregrine_par}), at $x=-25$).}
\label{fig5}
\end{figure}
\end{center}

\begin{center}
\begin{figure}[h]
\begin{tabular}{cc}
\includegraphics[width=.48\textwidth]{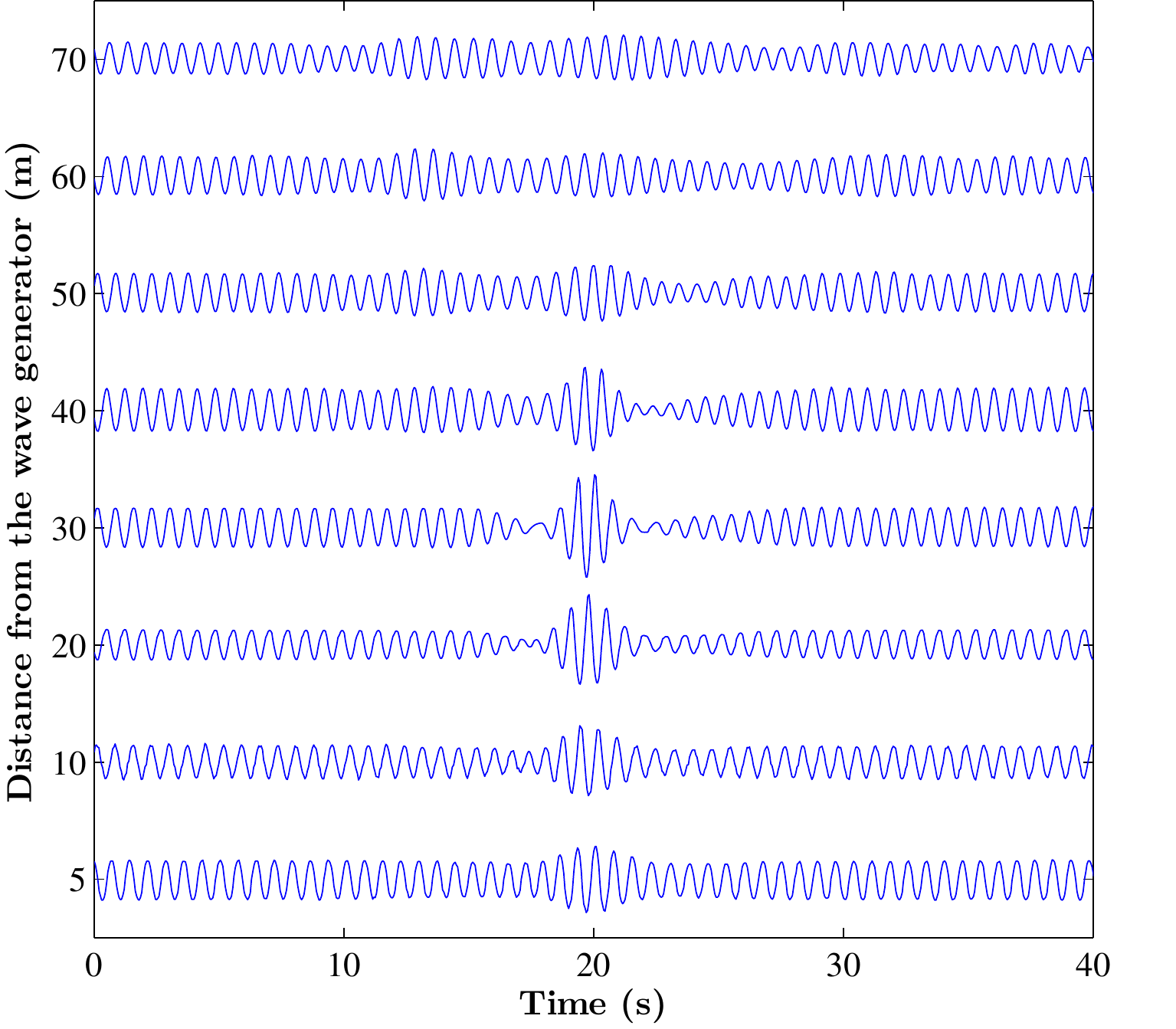}&
\includegraphics[width=.48\textwidth]{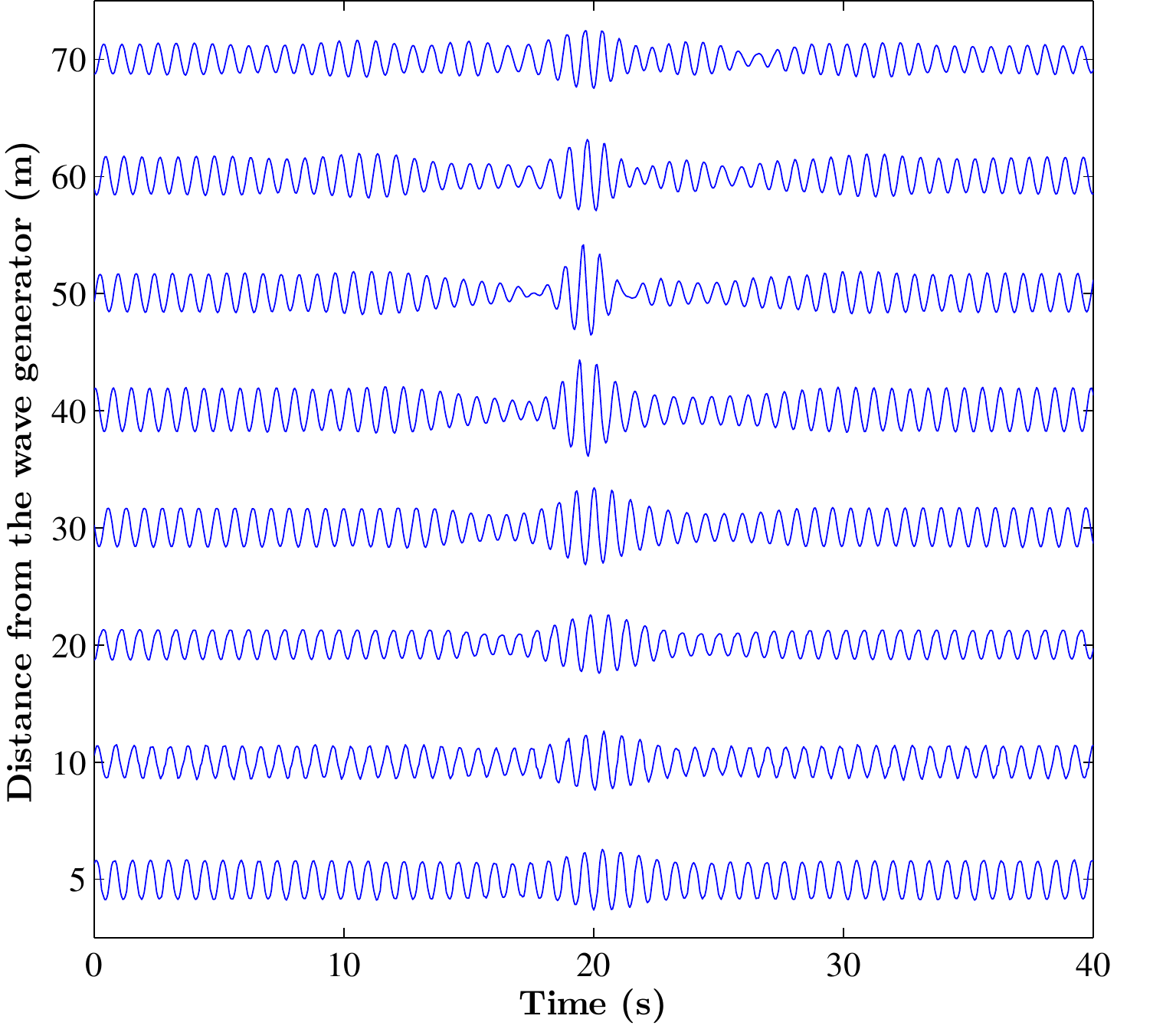}\\
\end{tabular}
\caption{Left: Propagation of Peregrine breather, starting with an exact initial condition, as described from exact Peregrine NLSE solution for $x=-25$ m. The amplitude of the carrier is $a=0.010$ m, while the steepness is $\varepsilon=0.07$. Right: Propagation of a similar initial breather structure for the same carrier parameters as in (Left), however with suppression of the characteristic Peregrine phase-shift imposed at $x=-25$ m in the initial conditions.}
\label{fig6}
\end{figure}
\end{center}

\begin{center}
\begin{figure}[h]
\begin{tabular}{cc}
\includegraphics[width=.48\columnwidth]{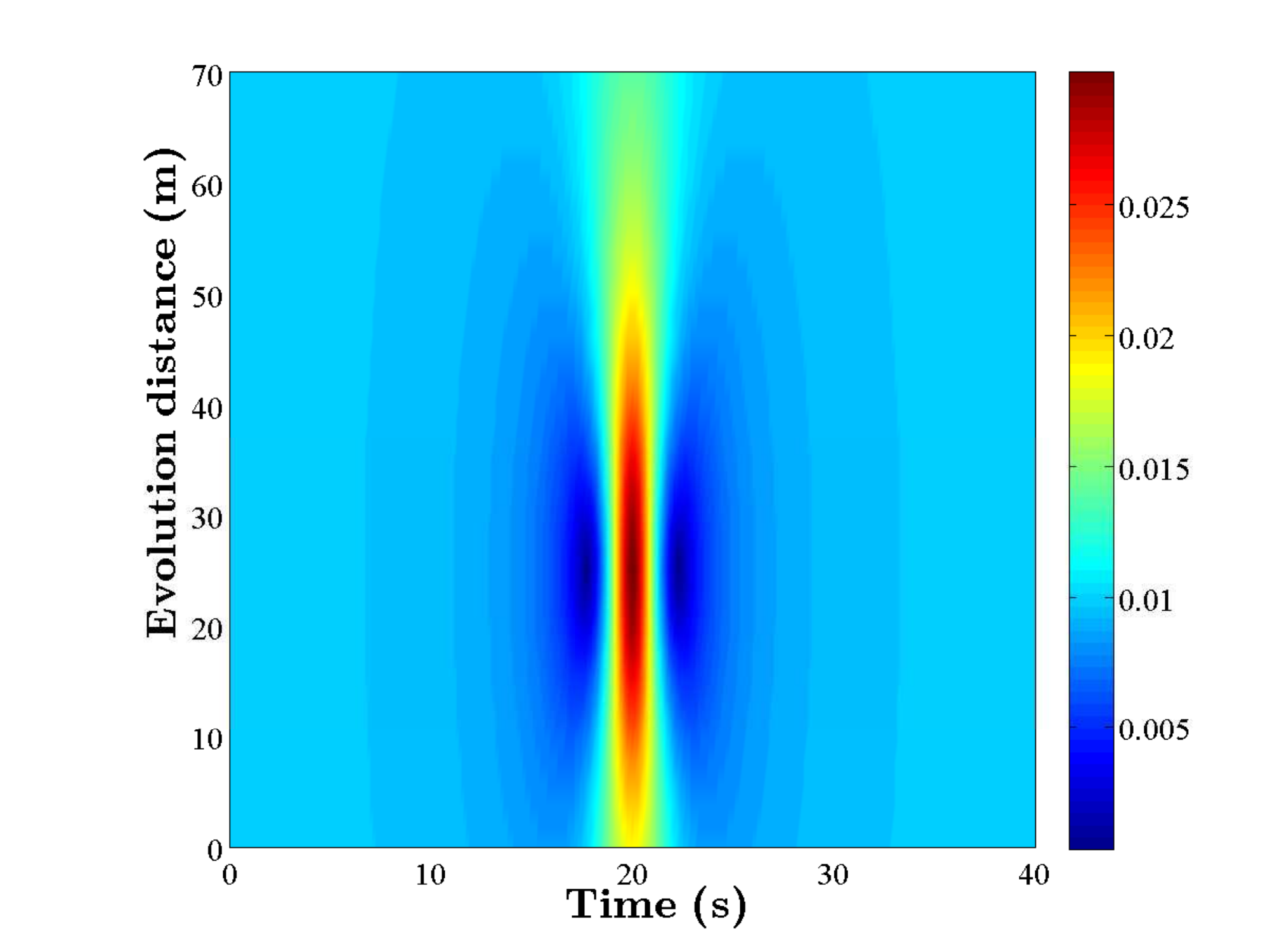}&
\includegraphics[width=.48\columnwidth]{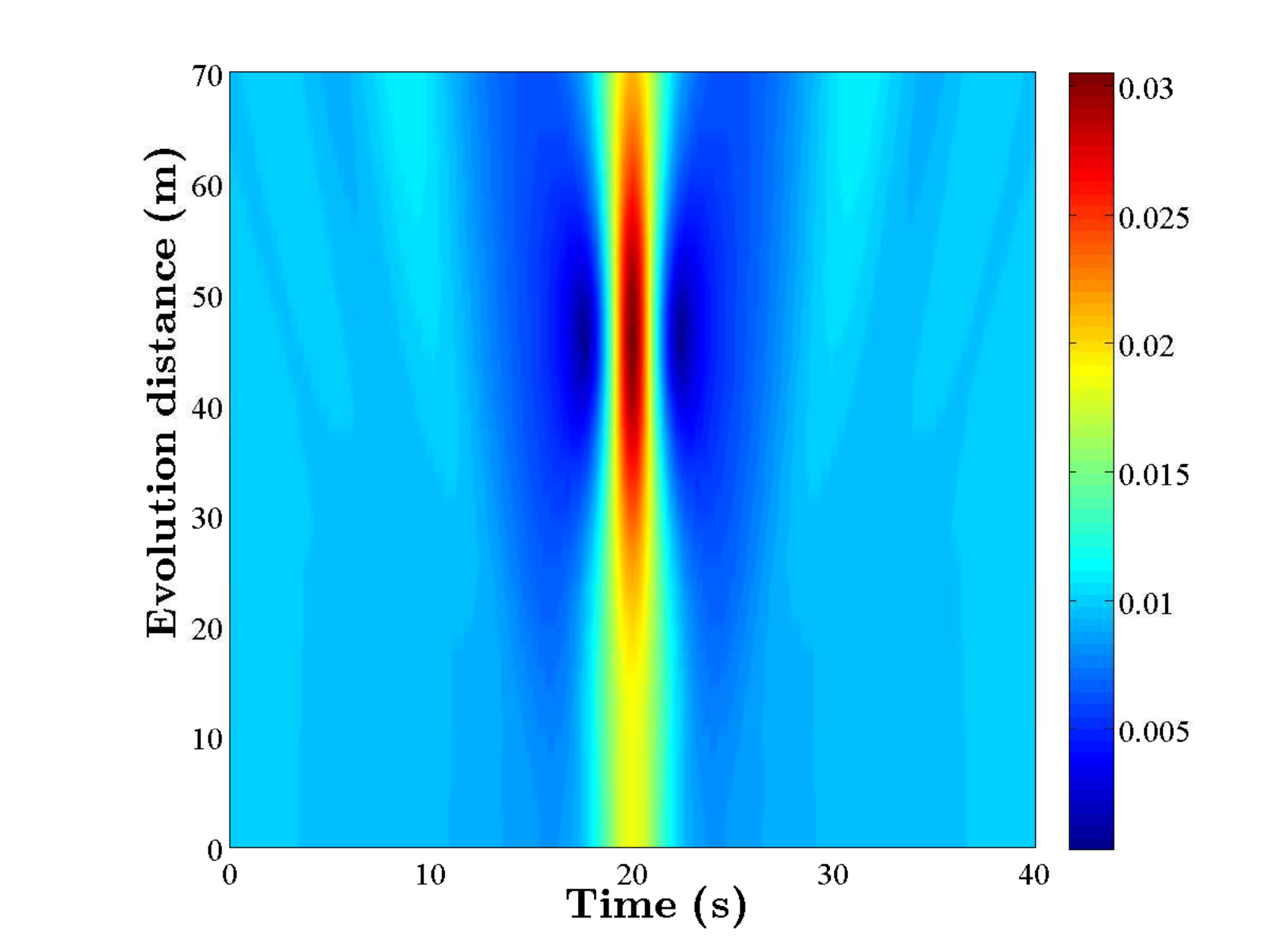}\\
\includegraphics[width=.48\columnwidth]{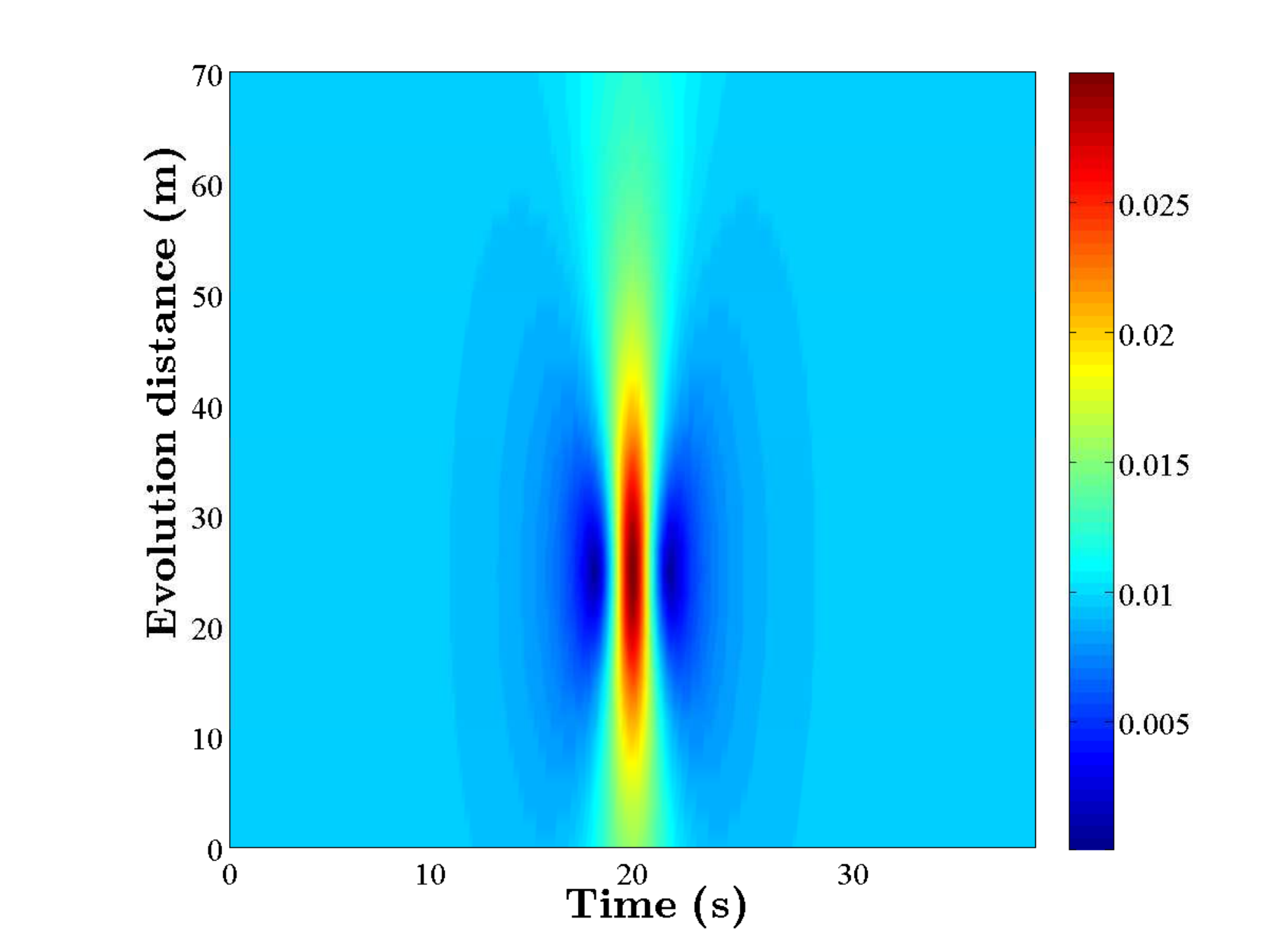}&
\includegraphics[width=.48\columnwidth]{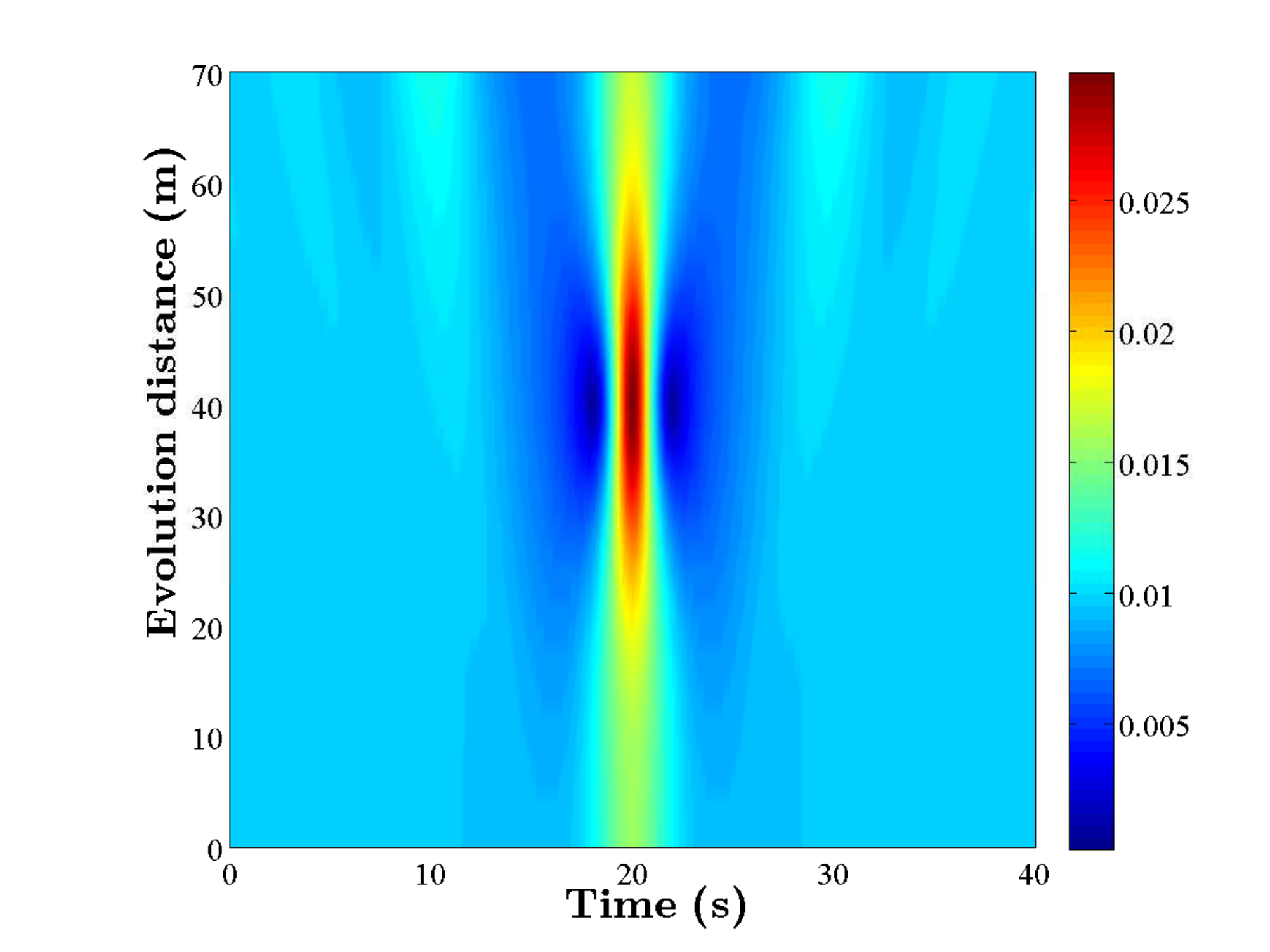}\\
\end{tabular}
\caption{
Numerical NLSE simulation of the experiments in Fig.~\ref{fig5}-\ref{fig6}. The top row is relative to the case shown in Fig.~\ref{fig5} (left: soliton input; right: suppressed phase input).
The bottom row is relative to the case shown in Fig.~\ref{fig6} (left: soliton input; right: suppressed phase input).
}
\label{fig7}
\end{figure}
\end{center}

As next, we discuss the experiments related to bright doubly-localized  (i.e., in $x$ and $t$) breather structures. Therefore, we will consider deep-water regime in the following. Due to the expected strong focusing of the wave field, we keep the steepness parameter small in order to expect a good agreement with weakly nonlinear theory. At the wave maker we excite either a plane wave with small envelope perturbation as given (in modulus and phase) by Eq. (\ref{Peregrine_par}) at $x=-25$, or the same type of wave, though with suppressed envelope phase. The results for the deep-water case for the carrier parameter $a=0.010$ m and $\varepsilon=0.06$ are depicted in Fig.~\ref{fig5}. 
Considering the significant propagation distance of 70 m, we can notice at first glance the growth and decay of the Peregrine breather. On the other hand, when the specific initial phase-shift at the input stage is ignored, a longitudinal retarded wave focusing dynamics is observed. Instead of a maximal focusing expected to occur after 25 m from the wave generator, it has been observed around 50 m. Note that, 
due to the finite number of wave gauges (namely, placed at spatial intervals of 10 m) and therefore the discrete character of water wave field measurements, 
we are not able to establish the exact position of maximal wave amplification. The role of initial phase manipulation in the propagation of an Akhmediev breather had been studied numerically \cite{erkintalo2011akhmediev}. This type of wave focusing retardation is indeed observed in our corresponding Peregrine breather experiments. We recall that the Peregrine breather is the limiting case of an Akhmediev breather when the modulation frequency tends to zero. In order to confirm these observations, we repeat the same type of experiment with the same carrier amplitude, however, for an increased wave steepness. This allows a faster evolution of the focusing process due to the increase of the nonlinearity of the wave field, compared to the previous case. 
Fig.~\ref{fig6} shows the results of measured wave profiles, assigned to these latter initial conditions of the experiment. 
The results in Fig.~\ref{fig6} confirm the same wave attributes, already noticed in the observations depicted in Fig.~\ref{fig5}. Namely, as the Peregrine breather evolves as expected according to the exact NLSE theory, that is, when exact initial conditions are injected to the wave maker, the evolution of the wave field when the initial Peregrine phases are not satisfied show a similar focusing feature in the evolution, however, noticeably retarded. In this latter case the maximal wave focusing occurs after 40 m and not 25 m from the wave maker. We also note a distortion of the wave field that may allow the follow-up focusing of ensuing wave packets. Following these tank observations, numerical NLSE simulations were performed in order to confirm these physical observations. The latter are shown in Fig.~\ref{fig7}. 
The numerical results in Fig.~\ref{fig7} effectively confirm the experimental observations and the dynamics the corresponding wave field undergoes, as shown in Figs. \ref{fig5} and \ref{fig6}. In fact, we can annotate the retardation of waves' maximal amplification. These simulations also allow to quantify the nature of the retardation of maximal wave amplification as well as the wave envelope distortion that result from the phase-shift prohibition. In the first case the spatial deviations for maximal wave focusing are of about 25 m whereas in the second of 15 m.   

\section{Phase evolution analysis}
\begin{center}
\begin{figure}[htb]
\begin{tabular}{cc}
\includegraphics[width=.48\columnwidth]{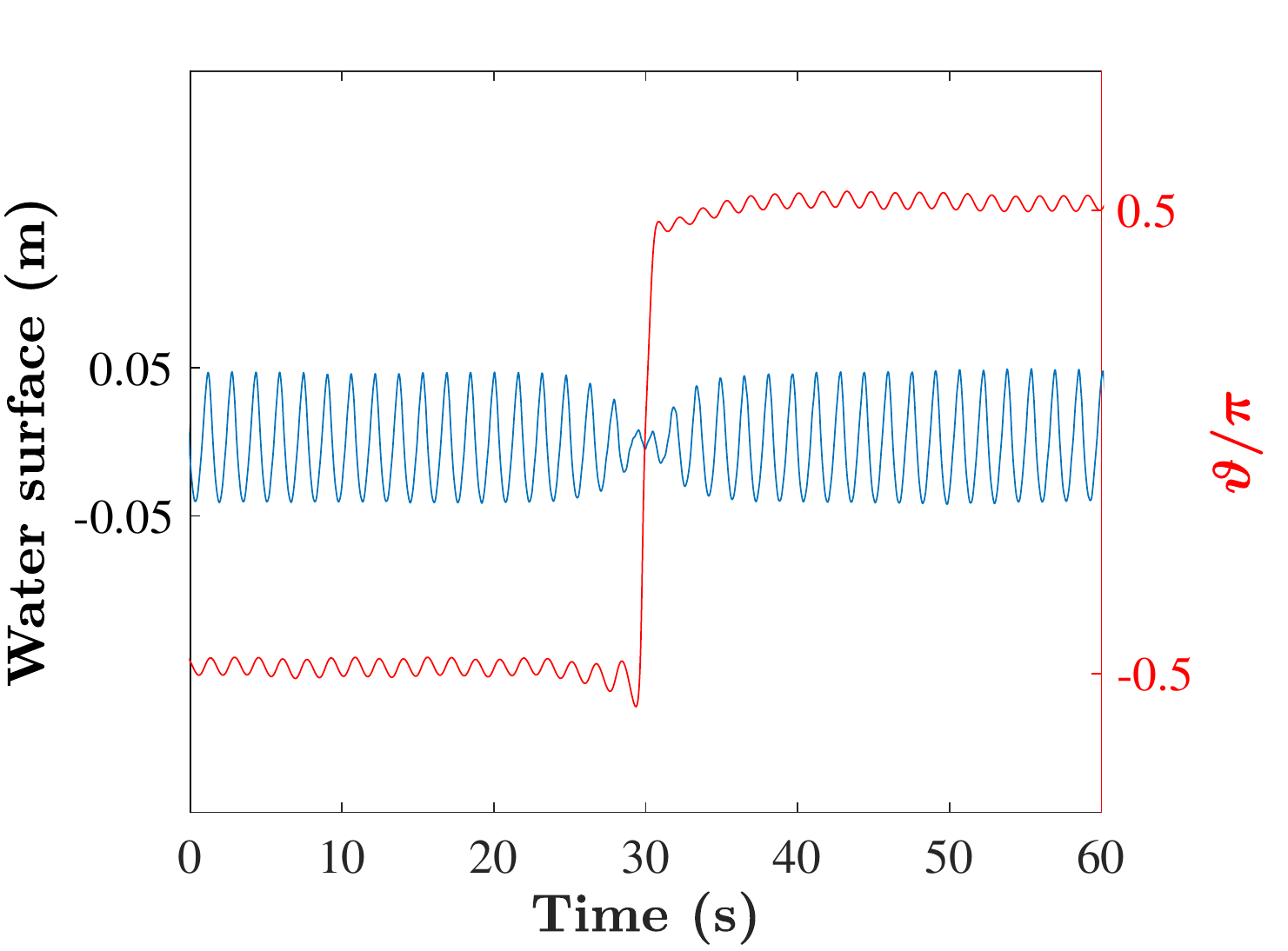}&
\includegraphics[width=.48\columnwidth]{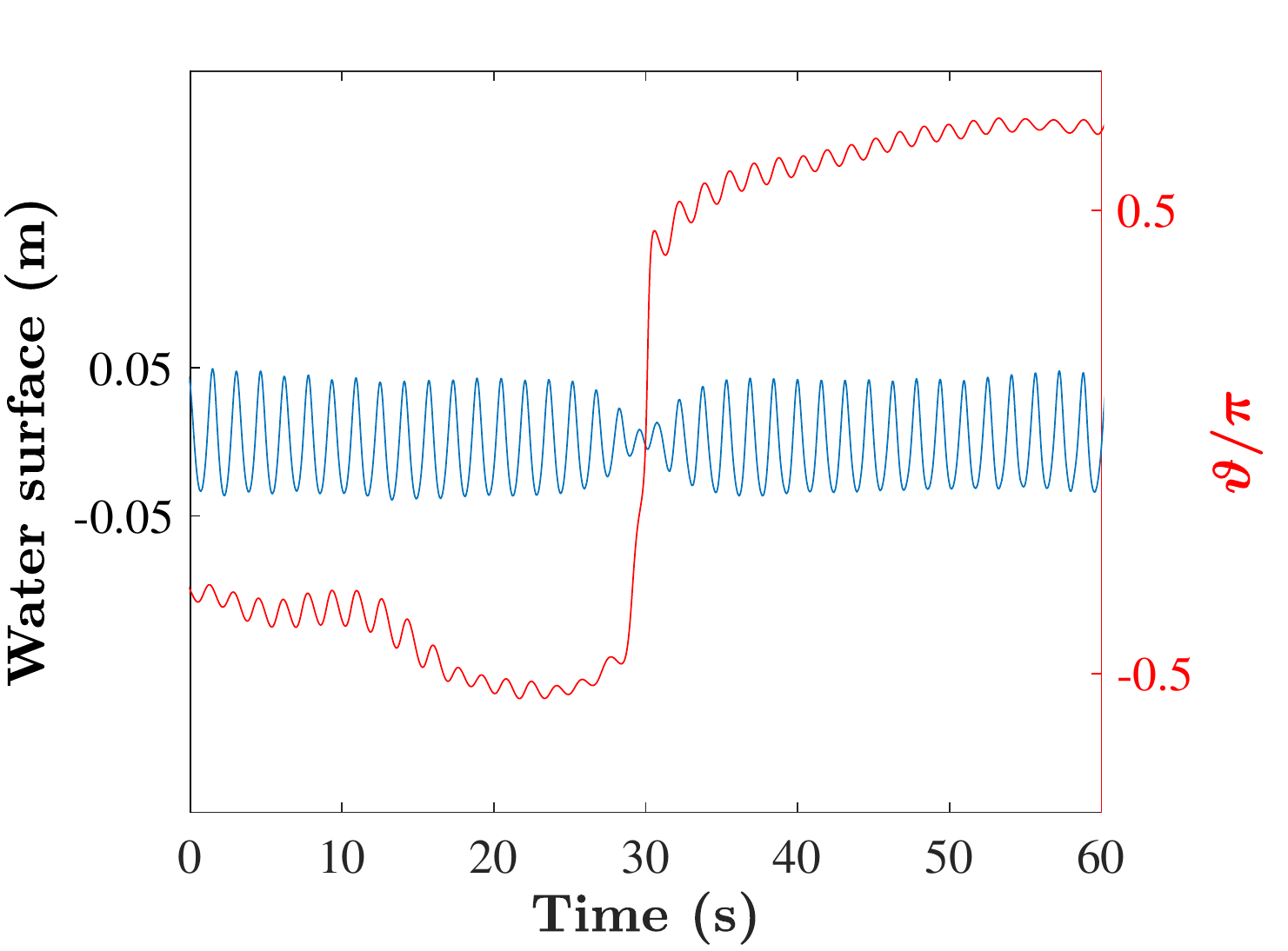}\\
\includegraphics[width=.48\columnwidth]{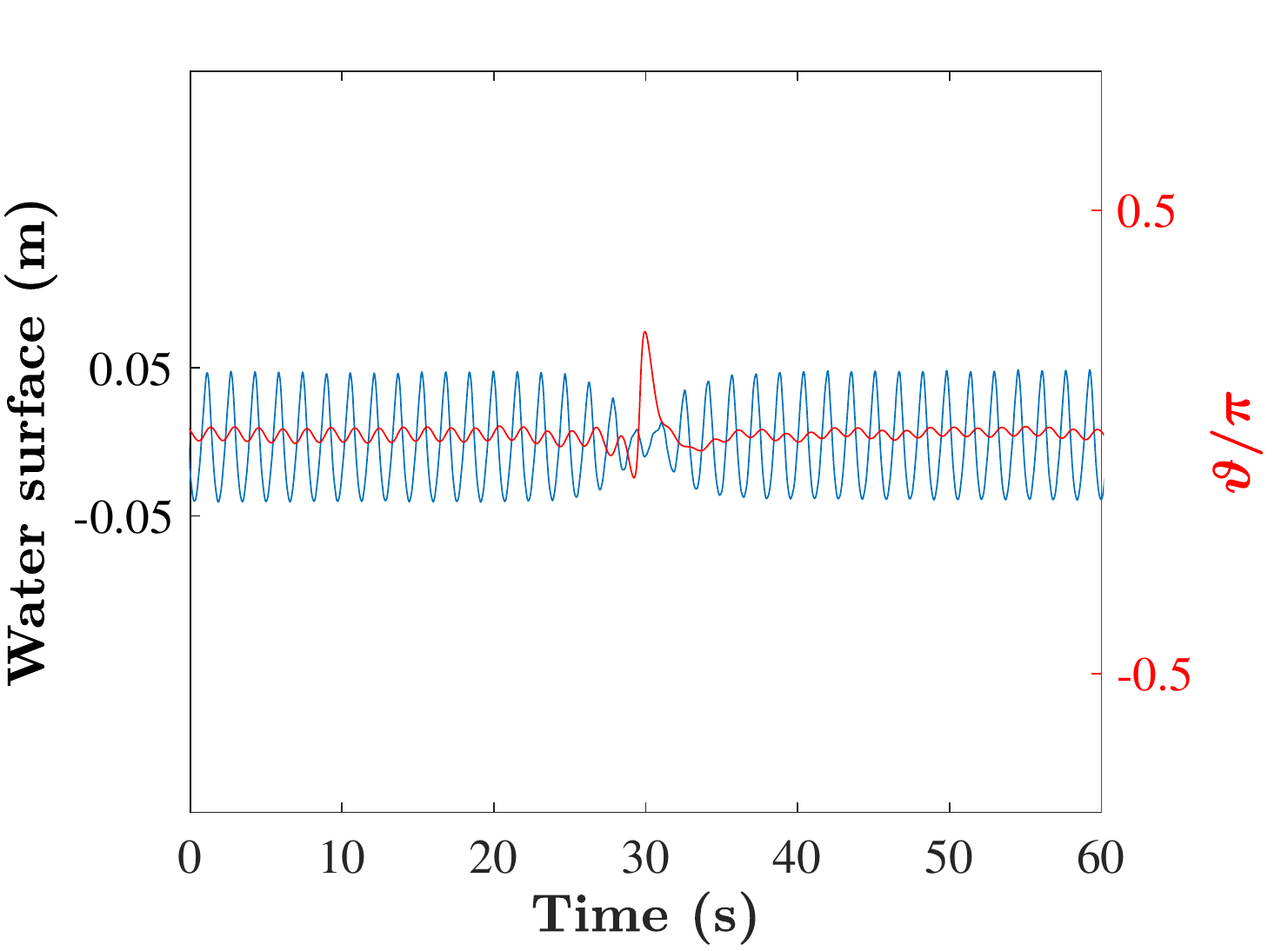}&
\includegraphics[width=.48\columnwidth]{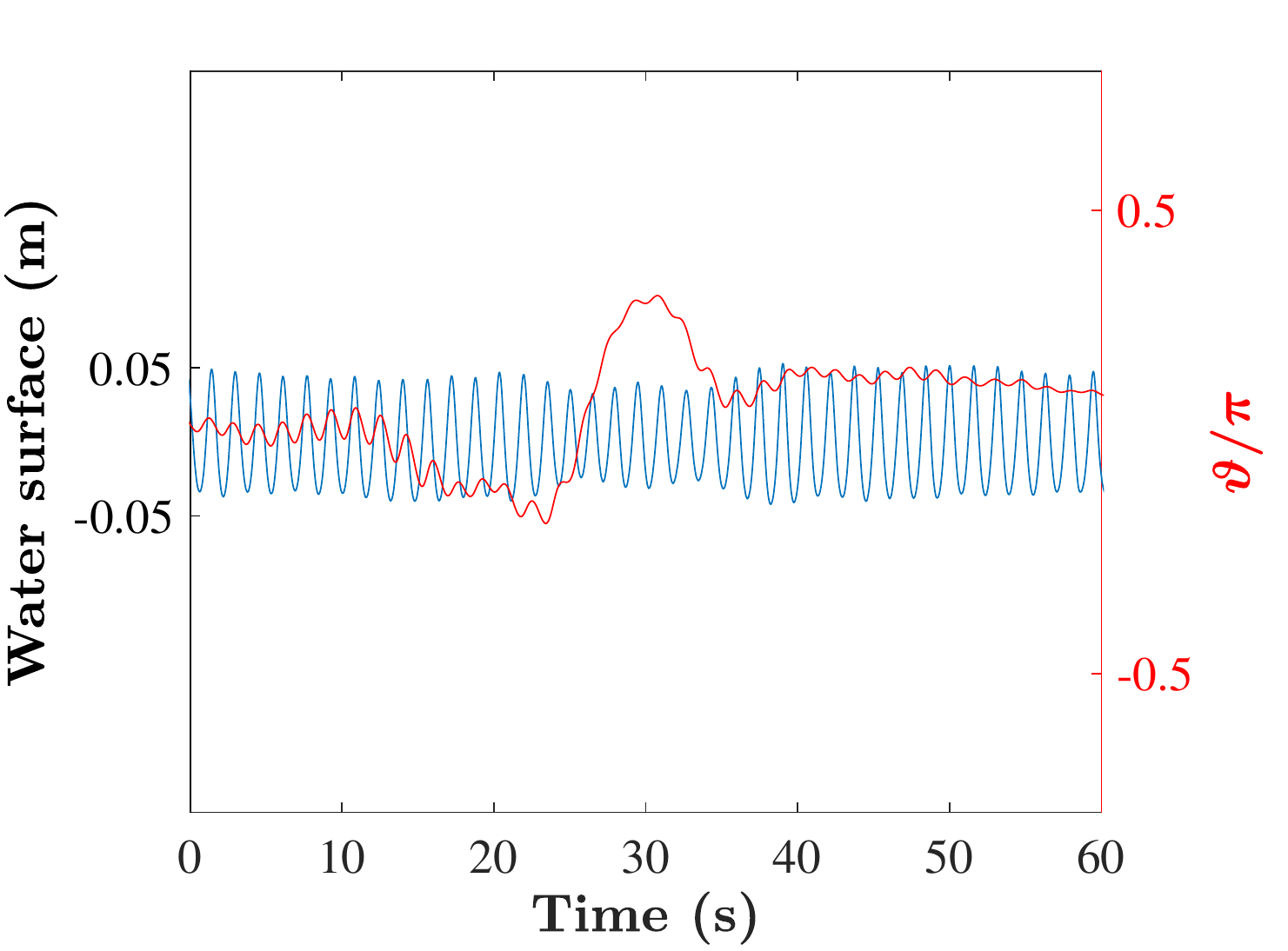}\\
\end{tabular}
\caption{Phase profiles generated by the input black envelope, as in the data set of Fig.~\ref{fig1}: blue and red lines display the measured profile of envelope amplitude and phase-shift, respectively, taken at the same longitudinal location. Top row, launch of exact black soliton (as in Fig.~\ref{fig1}, left panel). Here left and right panels refer to data from first gauge (5 m from wave generator) and last gauge (75 m from wave generator, respectively. Bottom row, launch of dark soliton with suppressed phase (as in Fig.~\ref{fig1}, right panel). As above, left and right panels refer to data from first gauge at 5 m, and last gauge  at 75 m, respectively. Here $\varepsilon=0.08$. 
}
\label{fig8}
\end{figure}
\end{center}

\begin{center}
\begin{figure}[htb]
\begin{tabular}{cc}
\includegraphics[width=.48\columnwidth]{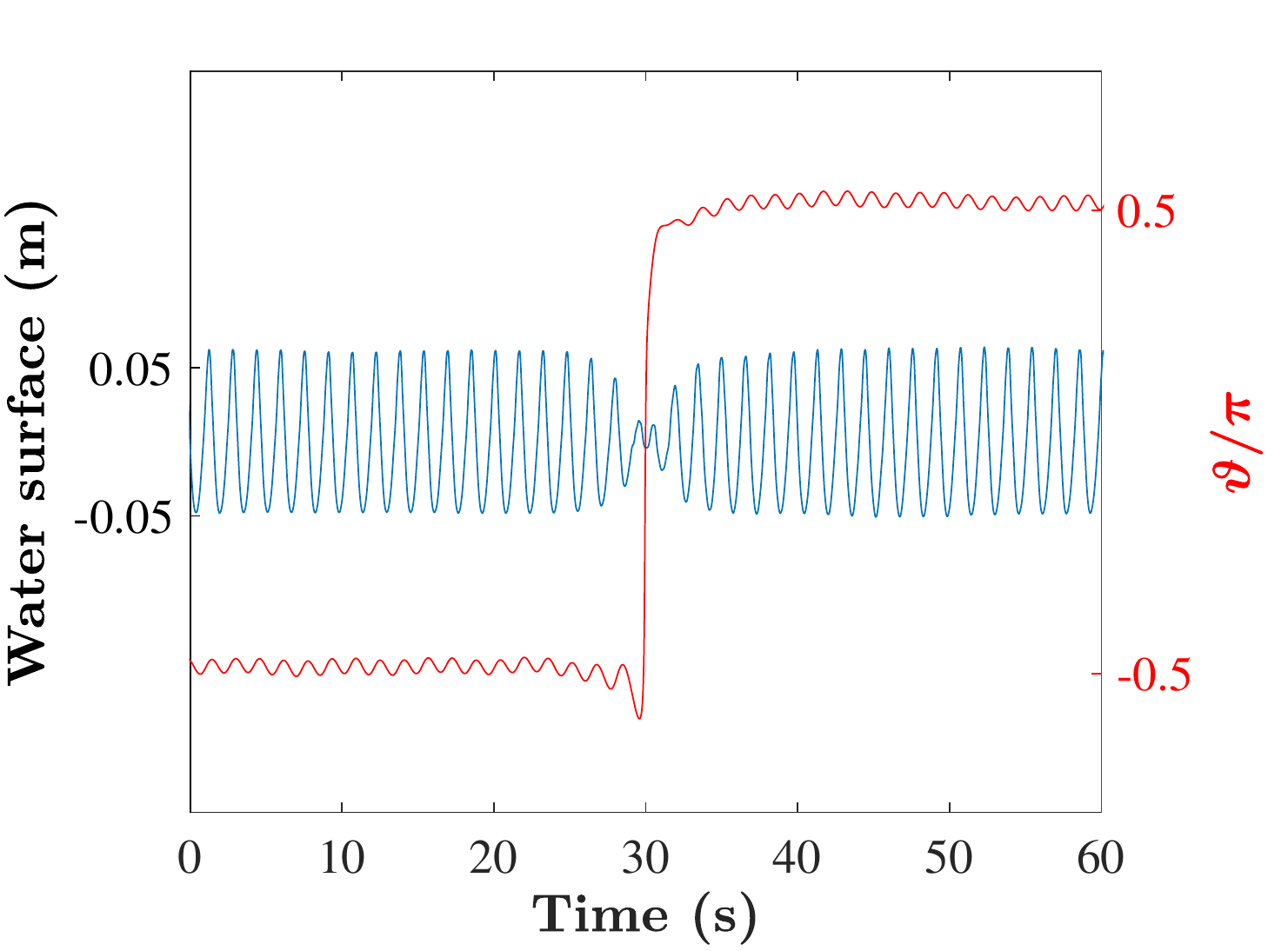}&
\includegraphics[width=.48\columnwidth]{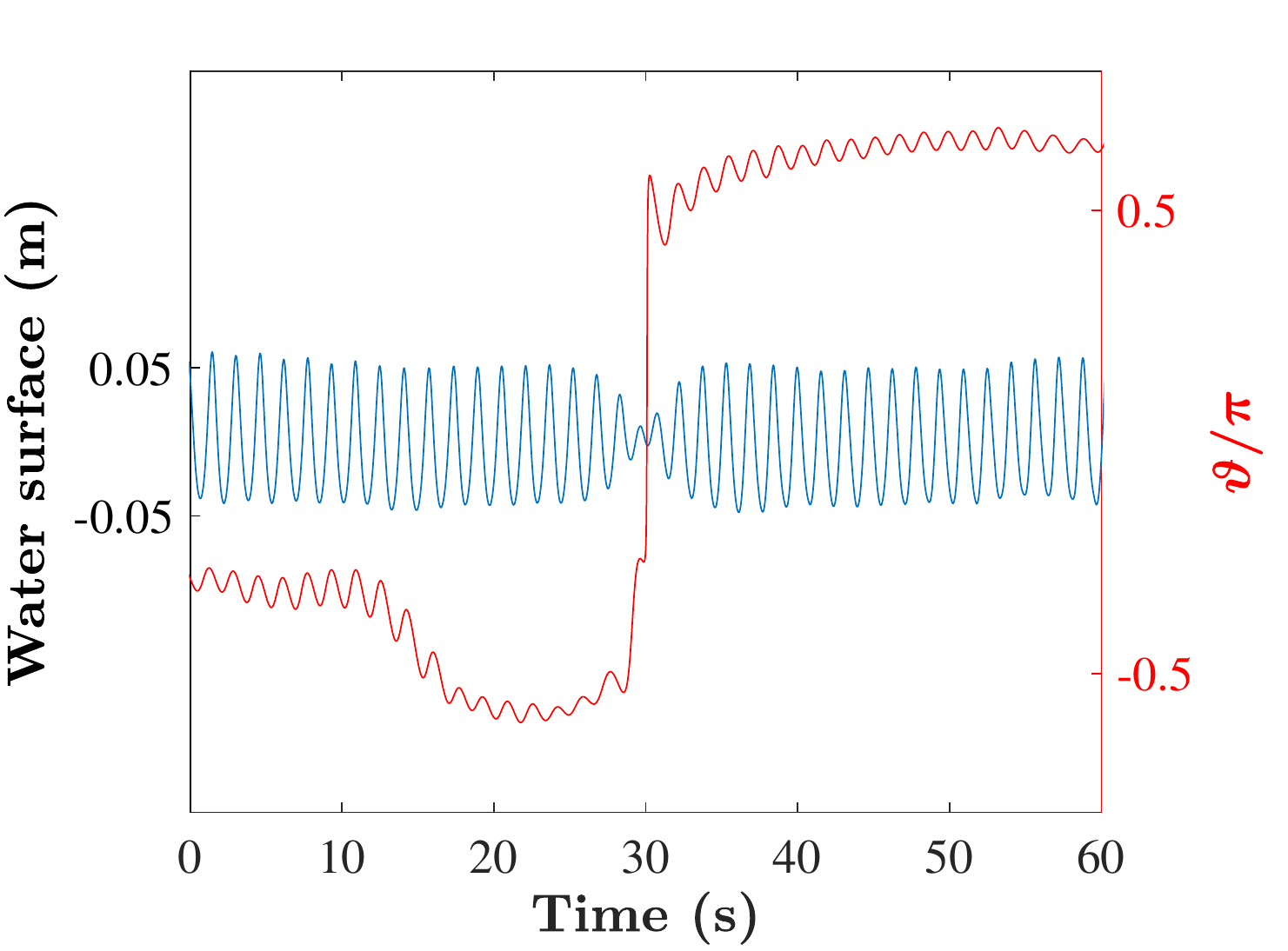}\\
\includegraphics[width=.48\columnwidth]{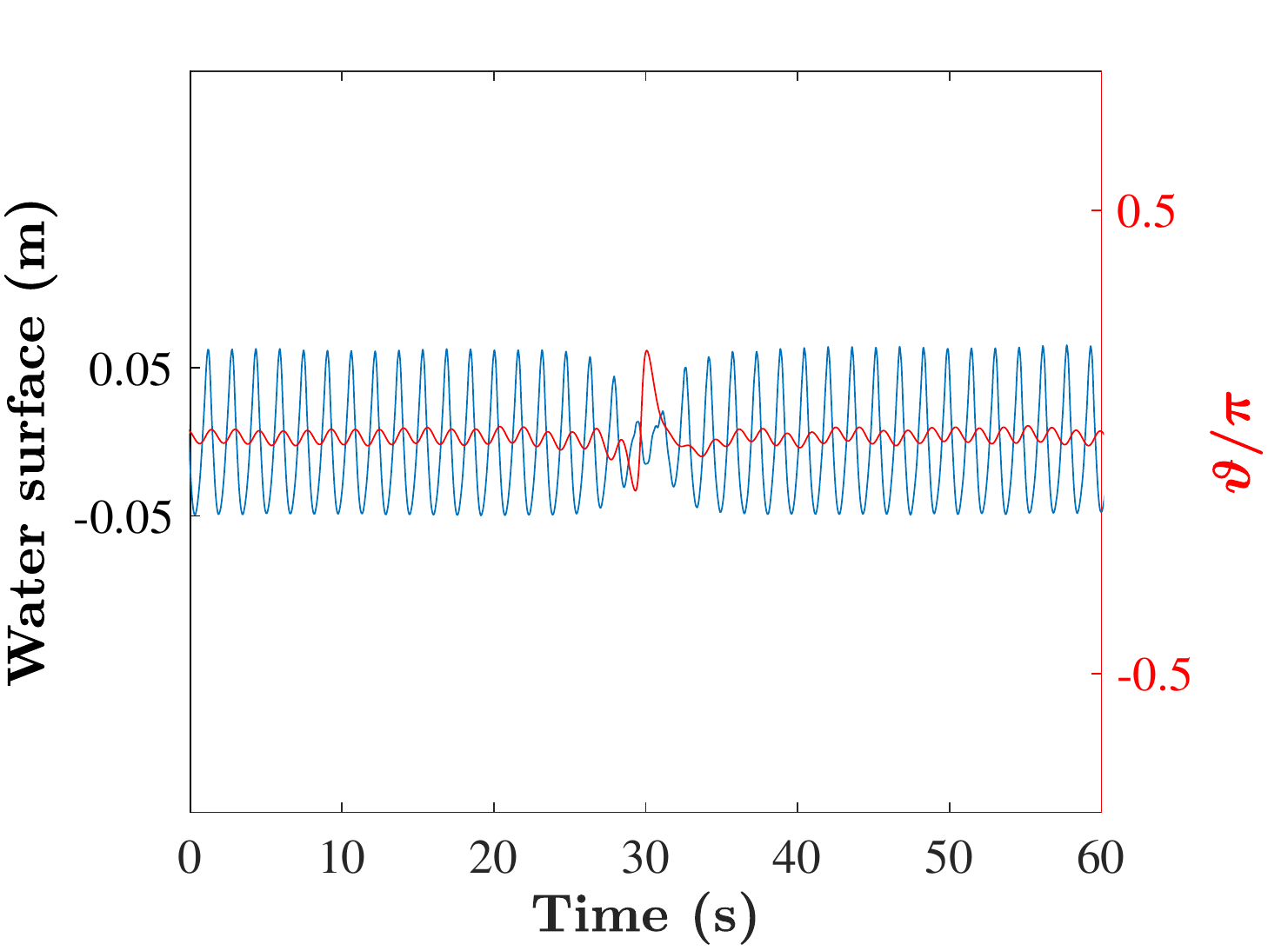}&
\includegraphics[width=.48\columnwidth]{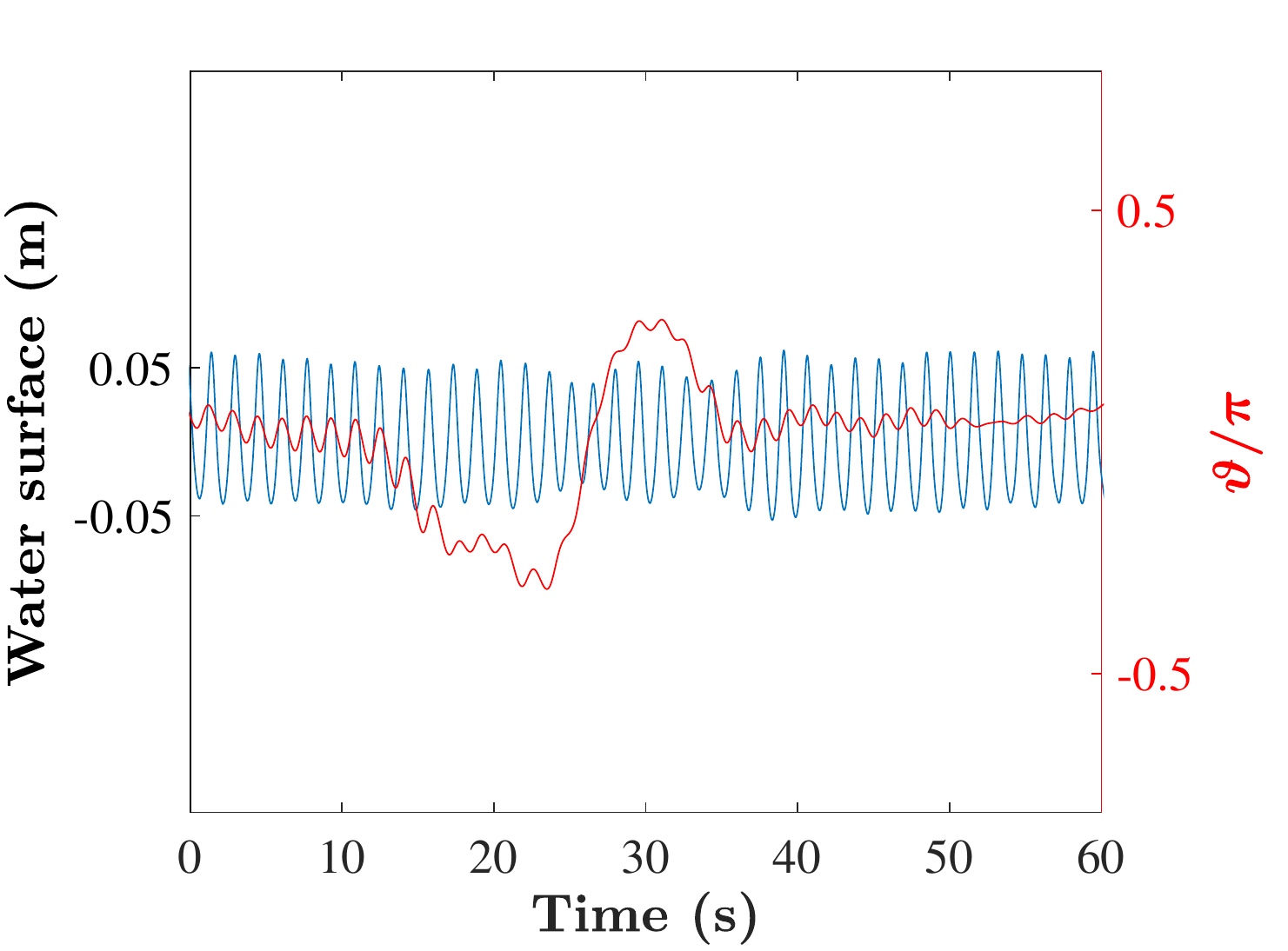}\\
\end{tabular}
\caption{
Same as in Fig.~\ref{fig8} for the data reported in Fig.~\ref{fig2} corresponding to a larger steepness $\varepsilon=0.1$.
}
\label{fig9}
\end{figure}
\end{center}

\begin{center}
\begin{figure}[h]
\begin{tabular}{cc}
\includegraphics[width=.48\columnwidth]{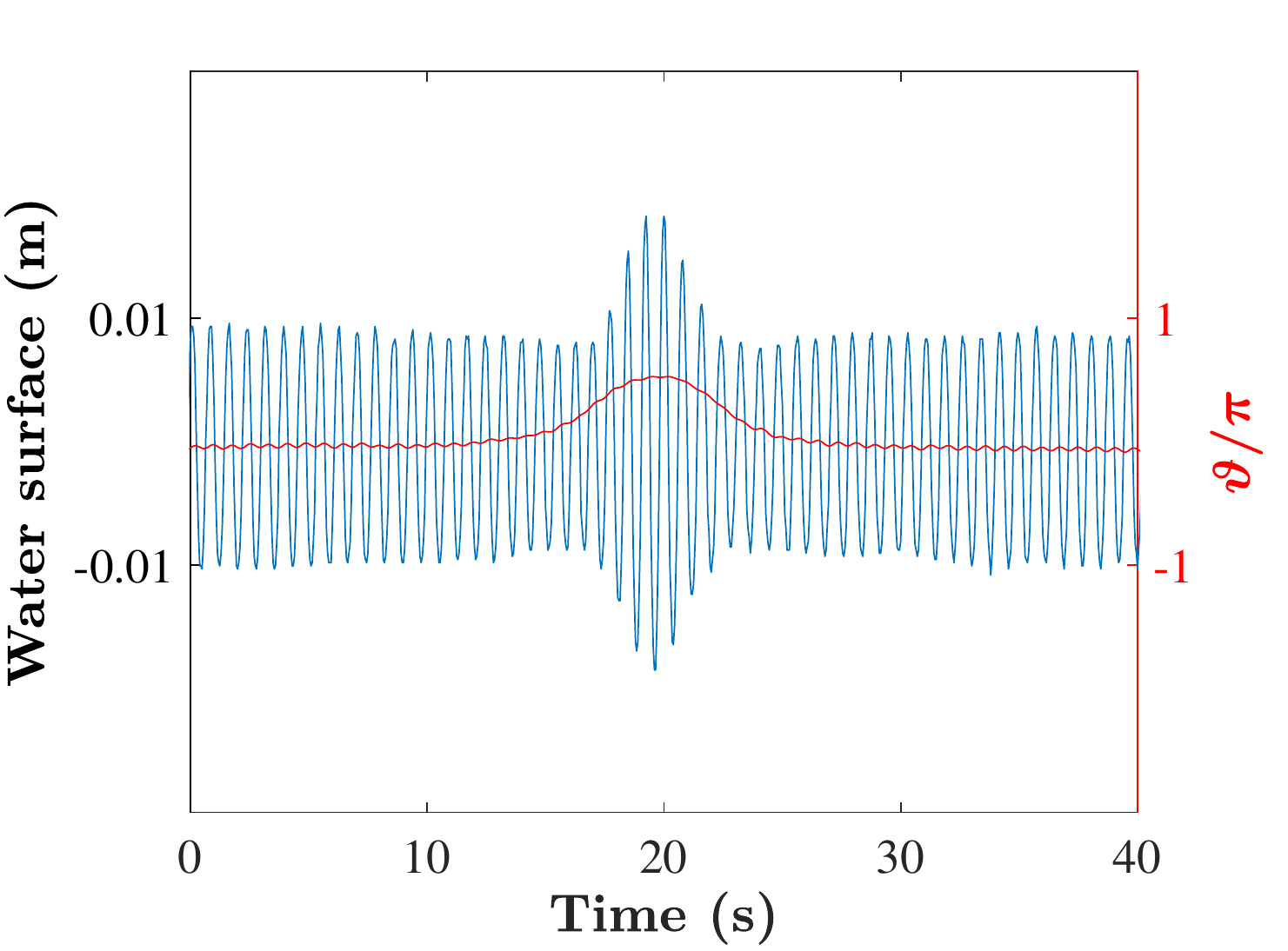}&
\includegraphics[width=.48\columnwidth]{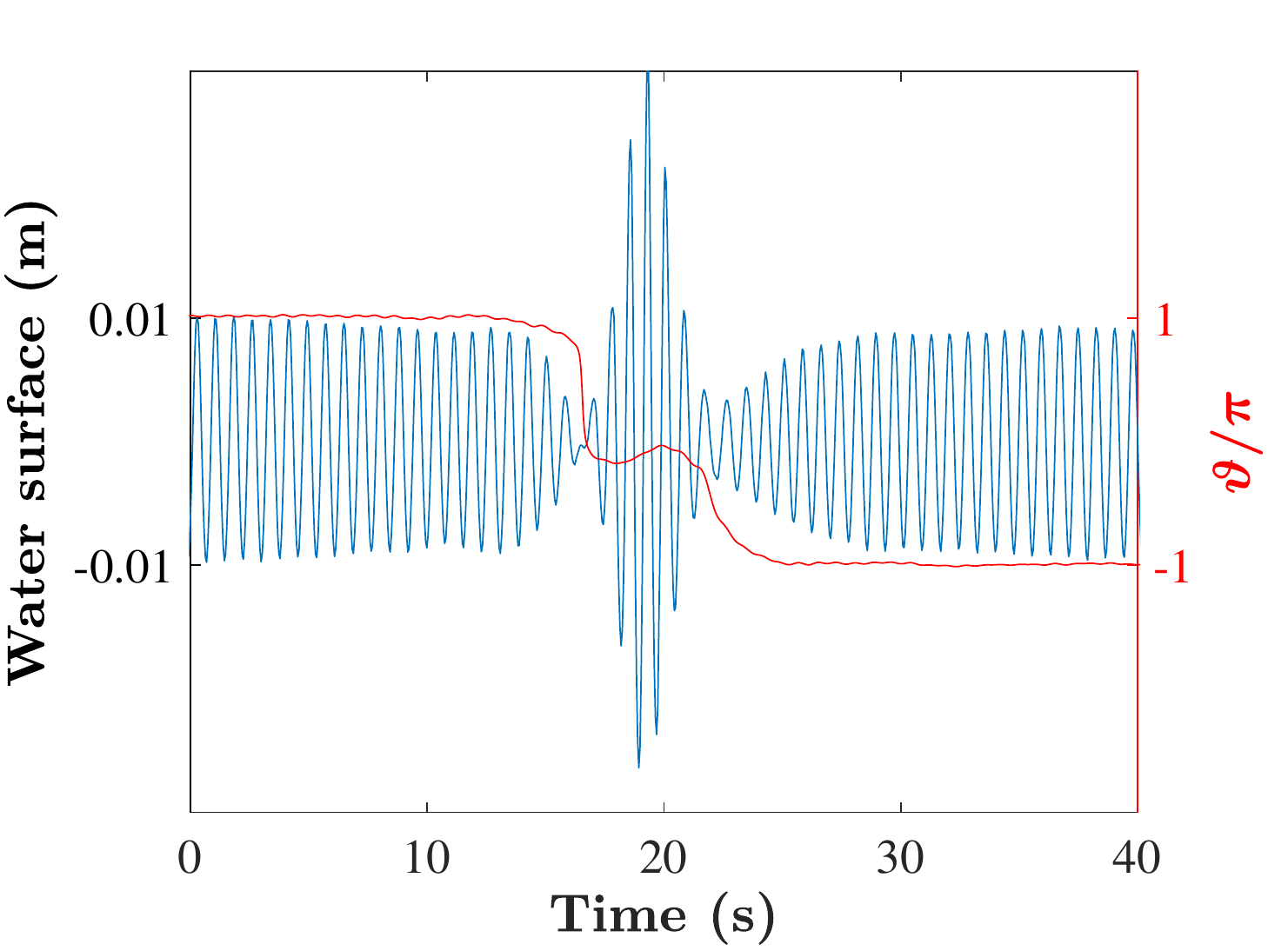}\\
\includegraphics[width=.48\columnwidth]{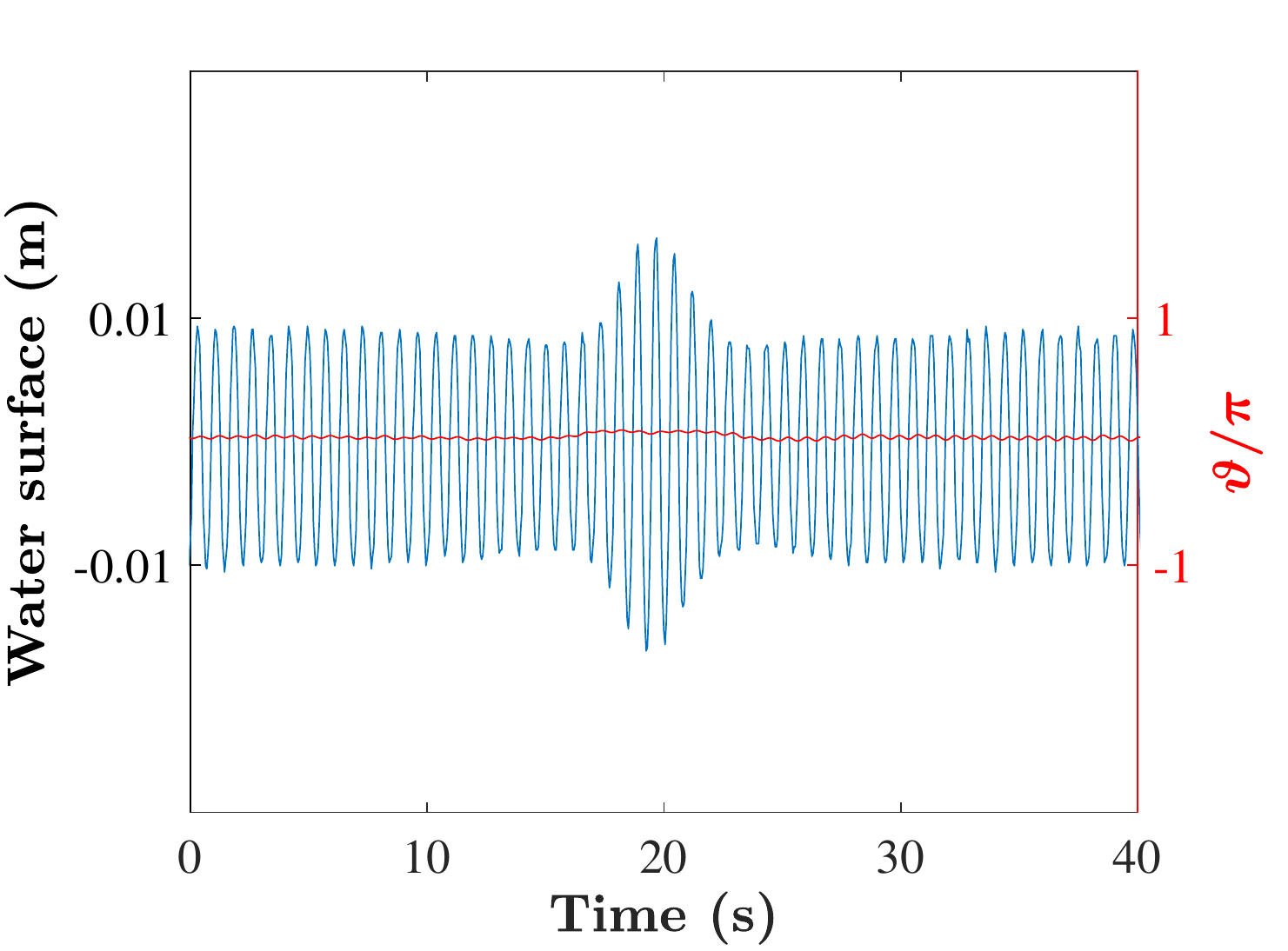}&
\includegraphics[width=.48\columnwidth]{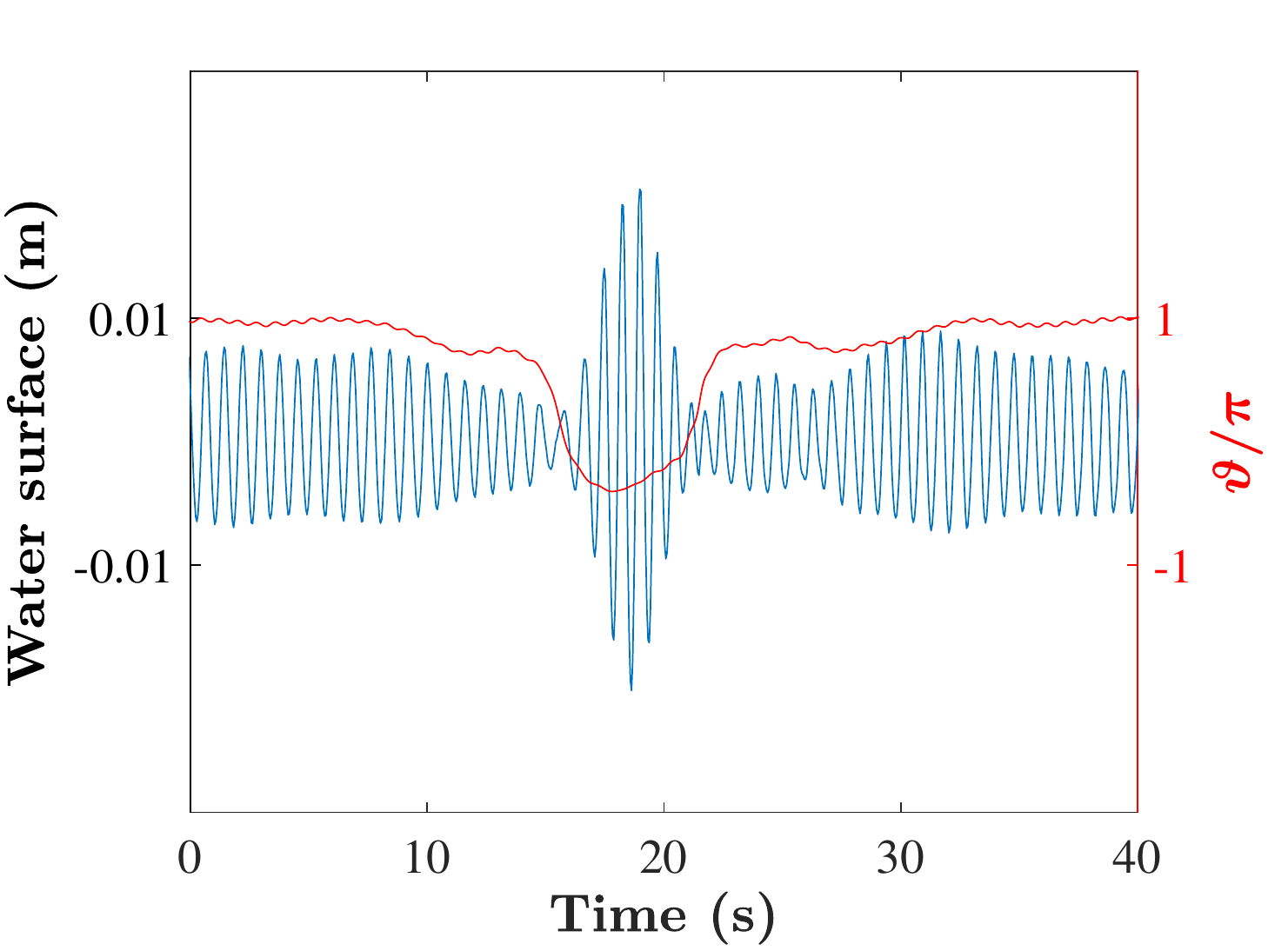}\\
\end{tabular}
\caption{
Phase profiles of the Peregrine breather waveform shown  in Fig.~\ref{fig5}: blue and red lines display the measured profile of envelope amplitude and phase-shift, respectively, taken at the same longitudinal location. Top row, launch of exact Peregrine (as in Fig.~\ref{fig5}, left panel). Here left and right panels refer to data from the first gauge ($5$ m from the wave generator) and the closest gauge to maximal amplification point of focusing ($30$ m from the wave generator). Bottom row, Peregrine with suppressed phase (as in Fig.~\ref{fig5}, right panel). As above, left and right panels refer to data from first gauge ($x=5$ m) and closest gauge to maximum amplification ($60$ m from wave generator). Here $\varepsilon=0.06$.
}
\label{fig10}
\end{figure}
\end{center}

\begin{center}
\begin{figure}[h]
\begin{tabular}{cc}
\includegraphics[width=.48\columnwidth]{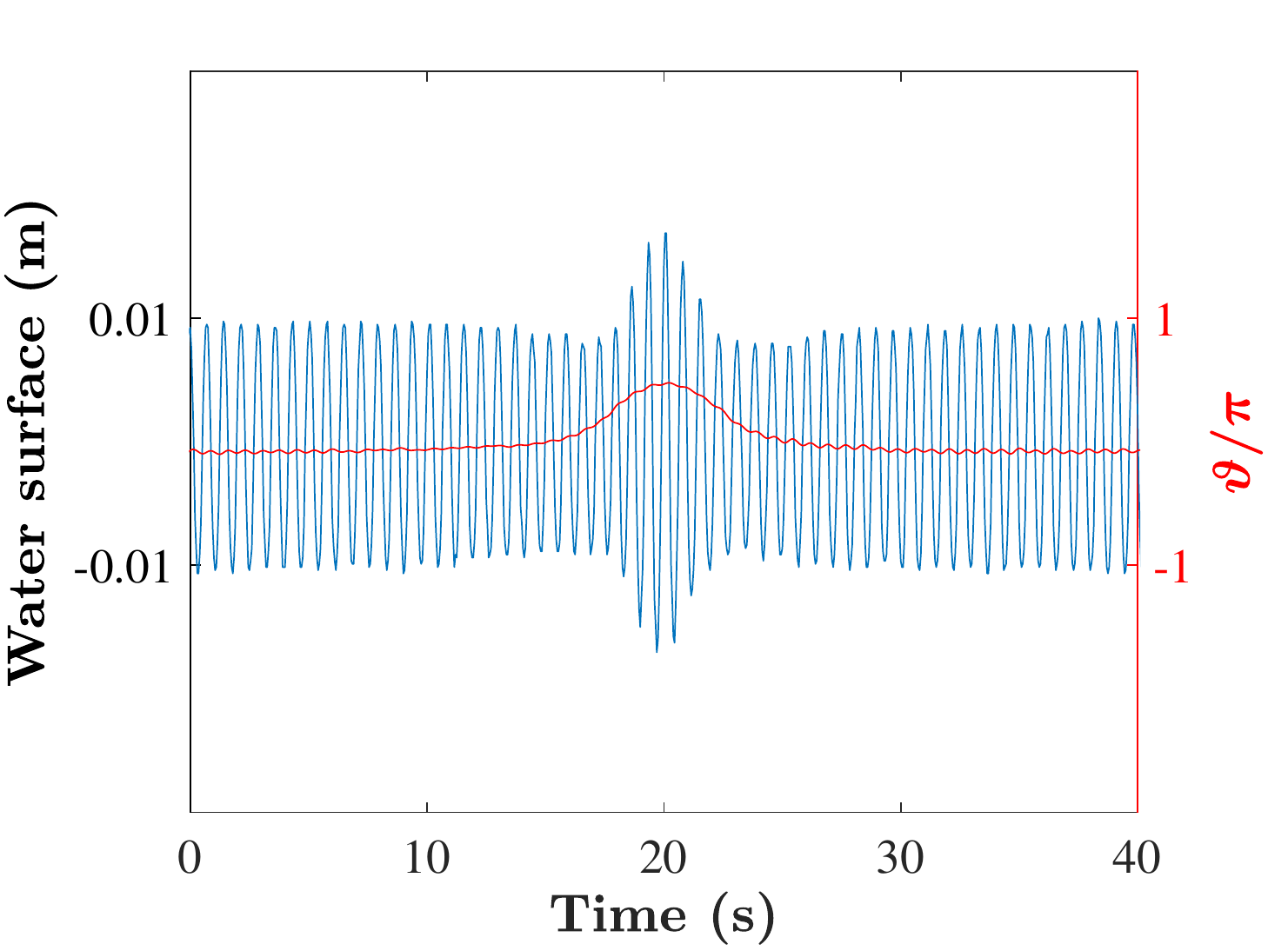}&
\includegraphics[width=.48\columnwidth]{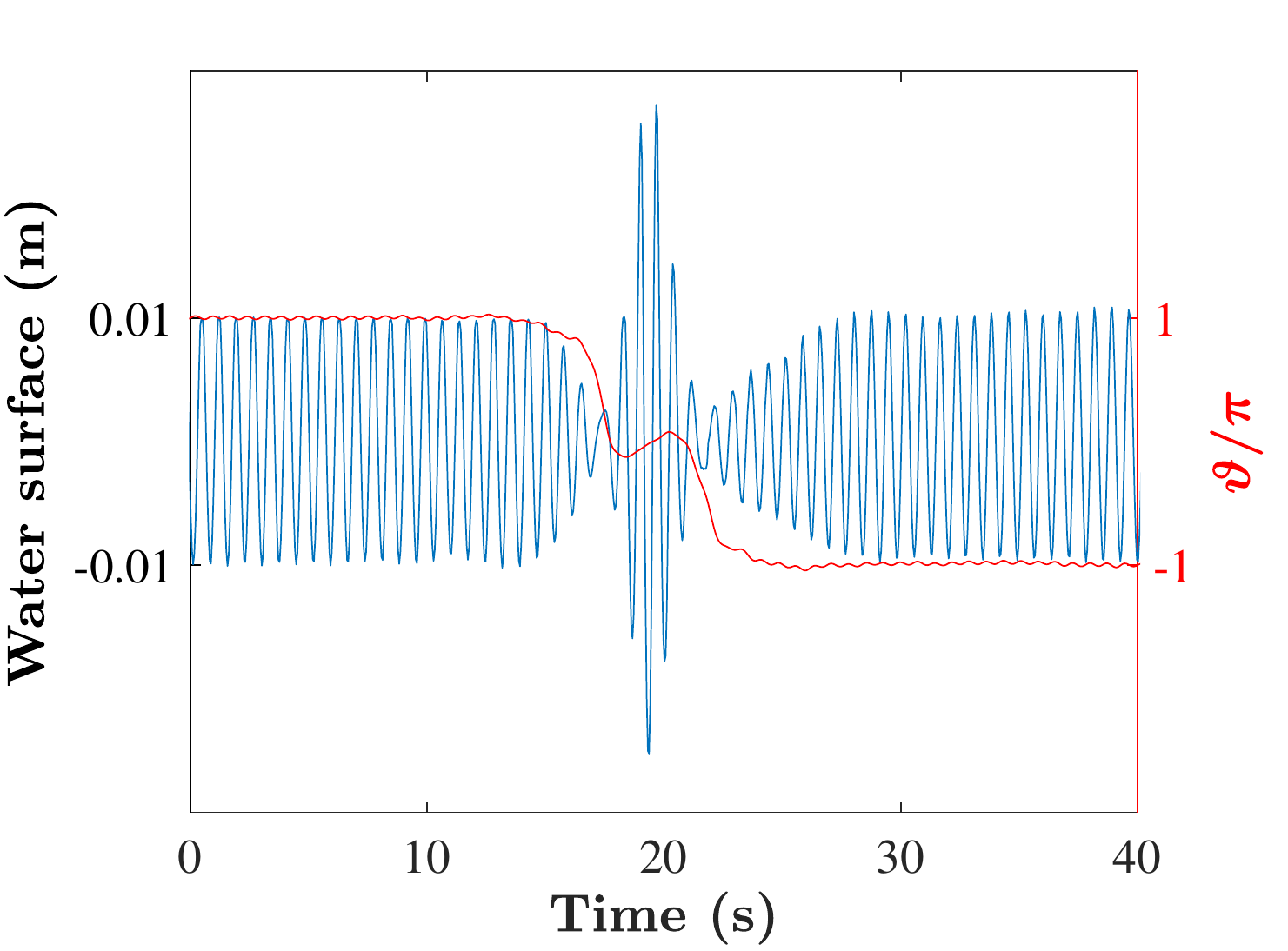}\\
\includegraphics[width=.48\columnwidth]{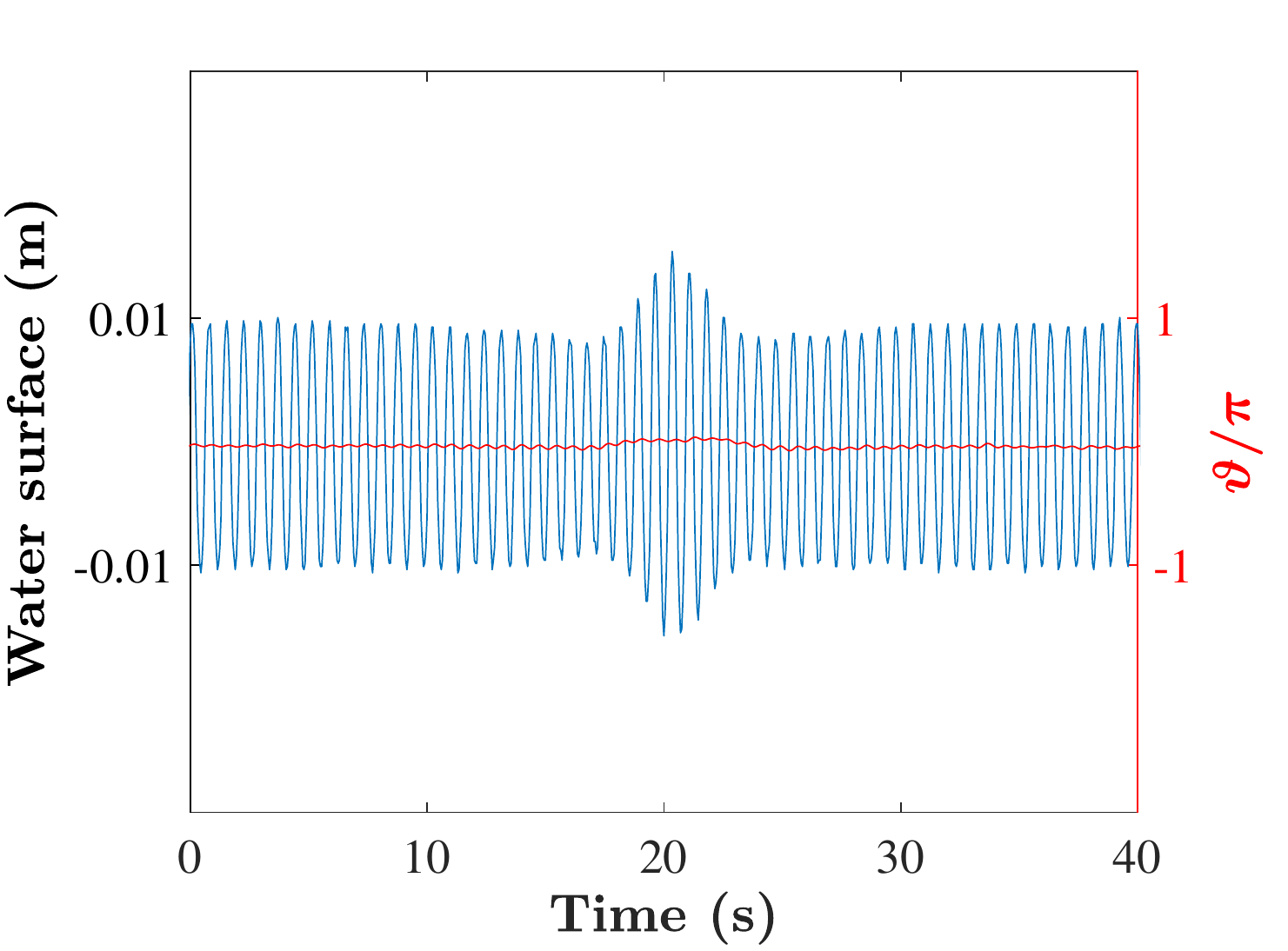}&
\includegraphics[width=.48\columnwidth]{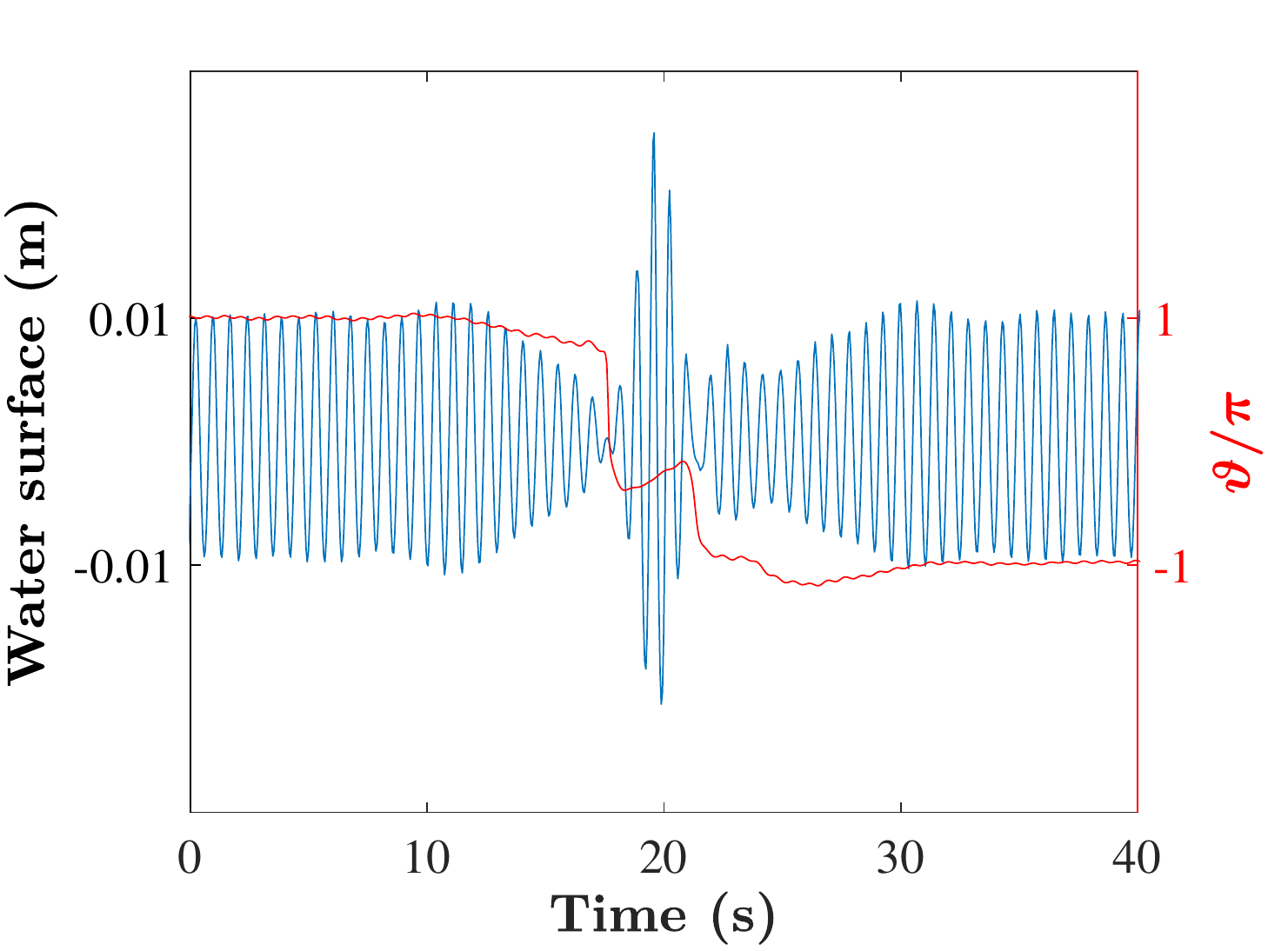}\\
\end{tabular}
\caption{Same as in Fig.~\ref{fig10} for the data reported in Fig.~\ref{fig6} corresponding to a larger steepness $\varepsilon=0.07$.}
\label{fig11}
\end{figure}
\end{center}

So far, a comparison of  the dark and the Peregrine solitons with the respective cases with equal envelope but phases set equal to zero has been done only in terms of the envelope amplitude. Here we make an extra effort and we compute the phases of the complex envelope from the time series of the carrier wave. The procedure is based on the theory of {\it analytic signals}  and relies on the following  procedure: using the inverse discrete Fourier transform, the Fourier amplitudes, $\hat\eta(\omega)$, of the time series  $\eta(t)$ of the surface elevation are numerically computed; then, positive frequencies are multiplied by 2 and negative ones by zero. The discrete Fourier transform is then used to calculate the filtered signal in physical space. The product of the original time series with the one obtained after filtering is the so-called analytic signal, $\eta_a(t)$; the complex envelope is then computed by removing the fast oscillation characteristic of the carrier wave as follows \cite{osborne2010nonlinear}:
\begin{equation}
\psi(t)=\eta_a(t) \exp[-\operatorname{i} \omega t],
\end{equation}
with $\omega$ the frequency of the carrier wave. The phases $\varphi(t)$ are then computed by standard means as 
\begin{equation}
\varphi(t)=\tan^{-1}\left(\frac{\Im[\psi(t)]}{\Re[\psi(t)]}\right).
\end{equation}
Such procedure has been applied to all our data sets and the results are displayed in Figs. \ref{fig8} and \ref{fig9} for the dark soliton and Figs. \ref{fig10} and \ref{fig11} for the Peregrine soliton, respectively. Note that this approach may not be applicable for complex and broad-banded temporal signals \cite{ma2010laboratory}.

In all Figs. \ref{fig8}-\ref{fig11} we superimpose the experimentally retrieved phase temporal profiles, reported as solid red lines, to the corresponding wave elevation (solid blue lines), obviously measured at the same longitudinal location.\\
Specifically, as far as the evolutions shown in Fig.~\ref{fig1} are concerned, we compare in Fig.~\ref{fig8} the phase profiles for the case of ideal soliton excitation (see top row  in Fig.~\ref{fig8}) to the case of suppressed phase input (bottom row  in Fig.~\ref{fig8}). In particular, the exact dark soliton clearly exhibits a phase-shift of $\pi$ across the vanishing dip of the field. As shown, the soliton phase-shift remains nearly unchanged from the first gauge (see top left panel  in Fig.~\ref{fig8}) to the last one ($x=75$ m, see top right panel  in Fig.~\ref{fig8}), except for a slight distortion from the flat phase-shift profile of the tails at $x=75$, which is more evident at earlier times ($t<30$ s, top right panel).
Conversely, when the input phase-shift is suppressed, the wave packet develops an intrinsic phase dynamic, which is already noticeable at the first gauge at $x=5$ m (see bottom left panel  in Fig.~\ref{fig8}), and which evolves at the farthermost gauge at $x=75$ m in the profile shown in the bottom right panel  in Fig.~\ref{fig8}.
The latter shows a peak value of the phase-shift (compared with the tails) close to the expected value of $0.36 \pi$ from NLSE integration.
The phase-shift profile  in Fig.~\ref{fig8}, bottom right panel, has positive slope across the left depression envelope amplitude and negative slope across the right depression envelope amplitude, consistently with the phase profile of dark solitons which exhibit negative and positive group velocity deviation (with respect to natural group velocity), respectively. 
A distortion of the profile (dip in the envelope phase for $t<30$ s in bottom right panel) is also present in this case, similar to the black soliton case (top right panel  in Fig.~\ref{fig8}).
Finally, we observe a similar scenario for the phase-shift, also for the larger steepness $\varepsilon=0.1$, as explicitly displayed in Fig.~\ref{fig9}, which correspond to the data set reported in Fig.~\ref{fig2}. We remark that what we report here constitutes: (i) the first direct evidence of the phase dynamics of a hydrodynamic black soliton; (ii) the evidence for the phase dynamics of the fissioning wave packets, which in other areas, such as optics or Bose-Einstein condensation, remains a challenging issue due to the involved fast scales. 


In Fig.  \ref{fig10} a similar analysis is performed for the experimental data dealing with the Peregrine solutions. 
The top row in Fig. \ref{fig10} shows the Peregrine, close to the input (top left panel; 5 m from the wave generator), and at the focus point (top right panel; 
30 m from the wave generator). 
In particular, the smooth temporal profile of the phase at 5 m slowly evolves to end up forming, at 30 m, two $\pi$-phase-shifts, one of each side of the maximum of the envelope. Indeed, once the Peregrine solution has reached the maximum amplitude, the $\pi$-phase-shifts are visible around the zeros of the envelope amplitude (top right panel in Fig.  \ref{fig10}).
A slight asymmetry around the peak amplitude, which can be noticed at 
30 m is attributed to higher-order effects \cite{dysthe1979note,shemer2013peregrine}, which are likely to cause the formation of the exact zeros on the left and right side of the peak at slightly different propagation lengths (not detectable in the experiment due to discreteness of wave gauge positions). These discrepancies in large-amplitude wave profiles can be tamed by decreasing the wave steepness \cite{chabchoub2012experimental}. Nevertheless, here, the comparison with the phase-suppressed case in the input is even more interesting because,  as explicitly shown in Fig. \ref{fig10} comparing the top row and the bottom row, a $\pi$-phase-shift develops also in the case when the phases are set to zero in the initial condition, i.e. even if the phase of the perturbation of the background strongly deviates from that of the exact solution. 
This is a remarkable fact that finds its roots in the universality of the Peregrine soliton in the dynamics ruled by the NLSE \cite{bertola2013universality,tikan2017universality}; indeed,  in the limit of long perturbations (weak dispersion), it has been shown \cite{bertola2013universality} that the evolution of a wide class of bumps leads to the formation of a {\it local} Peregrine soliton. This is a rigorous result in the semiclassical regime, for which the reader is referred to \cite{bertola2013universality} for more details. Our results seem to indicate the validity of this argument also if we strongly deviates from the semiclassical regime dominated by the nonlinearity, consistently also with observations in \cite{tikan2017universality}.

We stress here that the evolution, once reached its maximum amplitude, is locally a Peregrine soliton, i.e. it has the same envelope shape as a Peregrine solution, it displays a $\pi$-phase-shift at the points where the envelope touches vanishing amplitude and its amplification factor is equal to three. 
Remarkably, a very similar behavior is also exhibited by the evolution characterized by a larger steepness ($\varepsilon=0.07$), as shown in Fig. \ref{fig11}.

Finally, we point out that the comparison between the left and right panels in the top rows of Figs. \ref{fig10}-\ref{fig11} show that the temporal phase profile of the Peregrine solution is not fixed but rather evolves during propagation, at variance with the phase of dark solitons. Indeed, at fixed time (e.g., at peak amplitude in time) one can define a spatially varying phase of the breather, which turns out to be a smooth function of x, and represents a local variation of the phase with respect to the background nonlinear phase \cite{Baronio2020radiation}. Such longitudinal phase profile, however, cannot be reliably determined from our experimental data, due to the limited number of gauges.

\section{Discussion and conclusion}
\begin{center}
\begin{figure}[h]
\begin{tabular}{cc}
\includegraphics[width=.49\columnwidth]{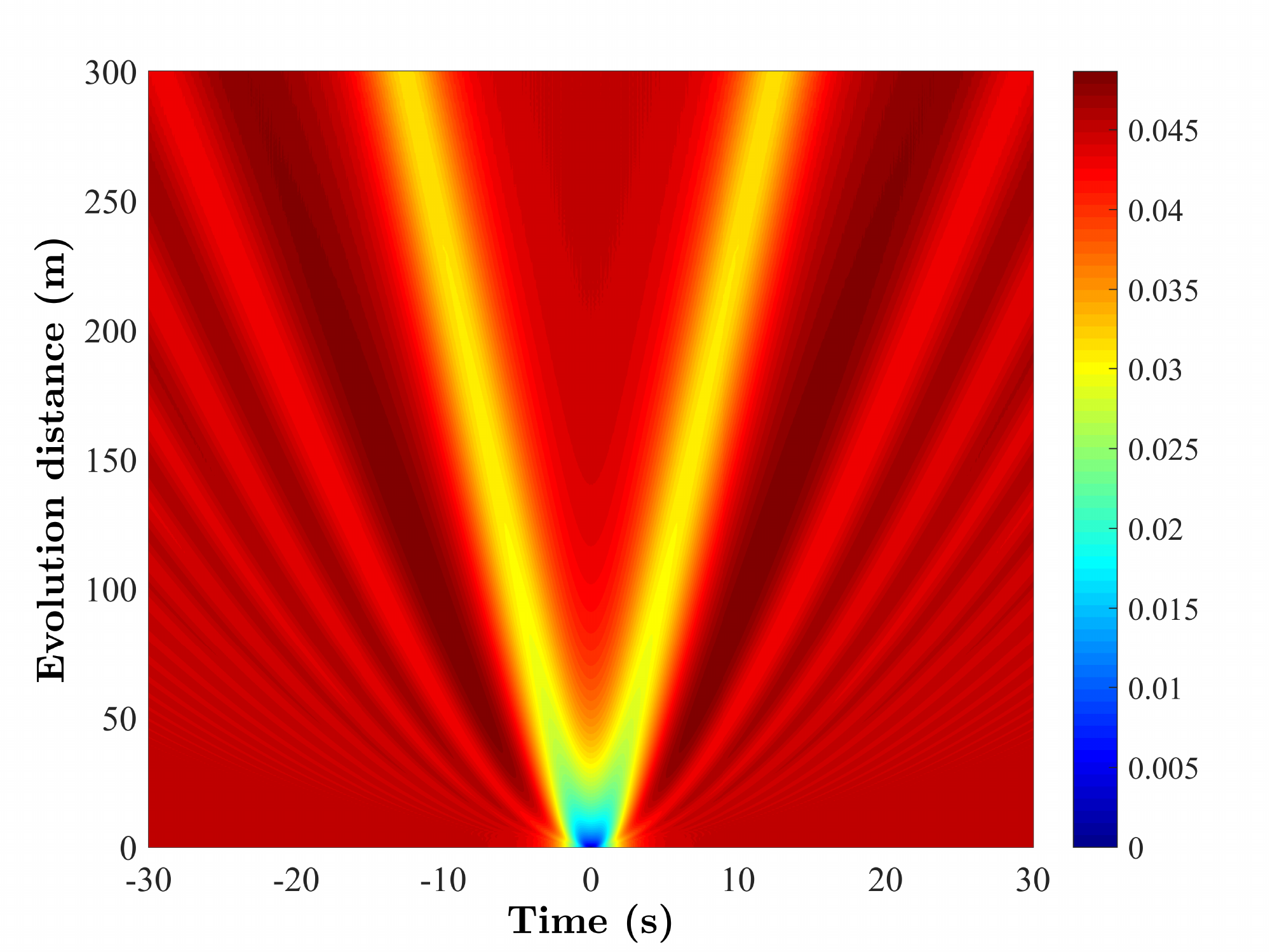}&
\includegraphics[width=.49\columnwidth]{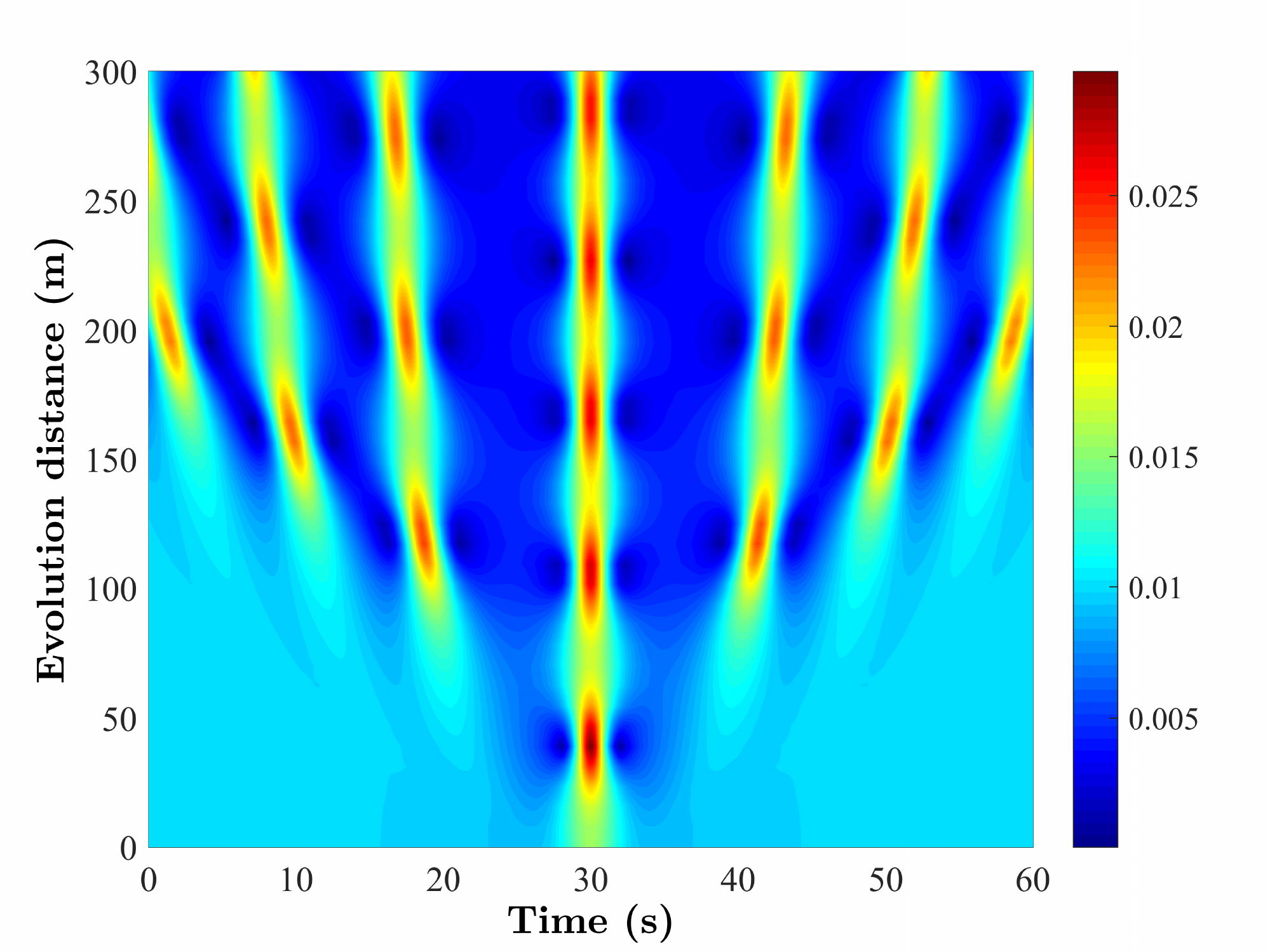}\\
\end{tabular}
\caption{Left: Propagation of the dark soliton envelope for carrier parameters as in Fig. 3 (bottom right panel) over a much larger propagation distance of 300 m. Right: Propagation of the Peregrine breather envelope for carrier parameters as in Fig. 7 (bottom right panel) over a much longer propagation distance of 300 m.
}
\label{fig12}
\end{figure}
\end{center}

To summarize, we have discussed experimentally and numerically importance of correct wave phase-shift settings in the boundary conditions for an experiment for the accurate propagation of localized NLSE envelope solutions, both of stationary- and pulsating-type in finite and infinite water depth, respectively. Two experimental campaigns were performed for different carrier parameters. Two unique large hydrodynamic wave facilities were used in this experimental study: the black soliton in finite water depth and the Peregrine model in deep-water. Both sets of experiments confirm the validity of weakly nonlinear NLSE theory, namely in the case of exact initial conditions that take into account the initial phase-shift and in the case when this information in the carrier wave is removed. In this latter case the initial localizations exhibit fission behavior into several localized structures of similar kind in very good agreement with our numerical simulations. To overcome the experimental restrictions that limit the distance for clean hydrodynamic propagation to 75 and 70 m, respectively, and hence do not allow to establish the nature of the asymptotic states, we performed further numerical simulations for the reported cases corresponding to the largest wave steepness values by increasing the distance to 300 m. These are shown in Fig.~\ref{fig12}. 
Indeed, we can observe an interesting type of fission expected from the two types of wave envelope models (black soliton and Peregrine breather). Both show similarities and differences in the fission behavior. 
Even though the basic features of the evolution dynamics can be interpreted as being dominated by dispersive effects only, yet for the dark soliton case we can see that 
the input localized dip effectively splits into two propagation-invariant gray envelopes with additional dispersive tails, in agreement with observations over 75 m. 
On the other hand, the Peregrine bright envelope shows, at distances substantially exceeding 70 m, a further break-up which produces a cascade of Peregrine type localizations that can be seen as limiting case of higher-order modulation instability \cite{erkintalo2011higher} or universal type of rogue wave cascade \cite{grimshaw2013rogue}, or, in a different language, the decay into a Kuznetsov-Ma soliton \cite{kuznetsov1977solitons,ma1979perturbed} coexisting with quasi-Kuznetsov-Ma symmetric pairs at non-zero velocities. We emphasize that further advanced hydrodynamic numerical simulations associated with inverse scattering analysis \cite{gelash2020anomalous} as well as motivated experiments in water and other nonlinear media would provide more information about the realistic distribution of such pattern. 

\bibliography{Refs}

\end{document}